\documentstyle[12pt]{article}

\catcode`\@=11
\long\def\@makefntext#1{
\protect\noindent \hbox to 3.2pt {\hskip-.9pt
$^{{\ninerm\@thefnmark}}$\hfil}#1\hfill}		

 \def\@makefnmark{\hbox to 0pt{$^{\@thefnmark}$\hss}}  

\def\ps@myheadings{\let\@mkboth\@gobbletwo
\def\@oddhead{\hbox{}
\rightmark\hfil\ninerm\thepage}
\def\@oddfoot{}\def\@evenhead{\ninerm\thepage\hfil
\leftmark\hbox{}}\def\@evenfoot{}
\def\sectionmark##1{}\def\subsectionmark##1{}}
 \def\beq{\begin{equation}}
\def\eeq{\end{equation}}
\def\beqn{\begin{eqnarray}}
\def\eeqn{\end{eqnarray}}
\def\bar{\begin{array}}
\def\ear{\end{array}}
\def\oe{\overline{e}}
\def\ou{\overline{u}}
\def\oc{\overline{c}}
\def\ot{\overline{t}}
\def\od{\overline{d}}
\def\os{\overline{s}}

\def\oh{\overline{h}}
\def\onu{\overline{\nu}}
\def\opsi{\overline{\Psi}}

\newcounter{sectionc}\newcounter{subsectionc}\newcounter{subsubsectionc}
\renewcommand{\section}[1] {\vspace{0.6cm}\addtocounter{sectionc}{1}
\setcounter{subsectionc}{0}\setcounter{subsubsectionc}{0}\noindent
	{\bf\thesectionc. #1}\par\vspace{0.4cm}}
\renewcommand{\subsection}[1] {\vspace{0.6cm}\addtocounter{subsectionc}{1}
	\setcounter{subsubsectionc}{0}\noindent
	{\it\thesectionc.\thesubsectionc. #1}\par\vspace{0.4cm}}
\renewcommand{\subsubsection}[1]
{\vspace{0.6cm}\addtocounter{subsubsectionc}{1}
	\noindent {\rm\thesectionc.\thesubsectionc.\thesubsubsectionc.
	#1}\par\vspace{0.4cm}}

\newcounter{appendixc}
\newcounter{subappendixc}[appendixc]
\newcounter{subsubappendixc}[subappendixc]

\renewcommand{\appendix}[1] {\vspace{0.6cm}
        \refstepcounter{appendixc}
        \setcounter{figure}{0}
        \setcounter{table}{0}
        \setcounter{equation}{0}
        \renewcommand{\thefigure}{\Alph{appendixc}.\arabic{figure}}
        \renewcommand{\thetable}{\Alph{appendixc}.\arabic{table}}
        \renewcommand{\theappendixc}{\Alph{appendixc}}
        \renewcommand{\theequation}{\Alph{appendixc}.\arabic{equation}}
        \noindent{\bf Appendix \theappendixc #1}\par\vspace{0.4cm}}

\def\abstracts#1{{
	\centering{\begin{minipage}{30pc}\tenrm\baselineskip=12pt\noindent
	\centerline{\tenrm ABSTRACT}\vspace{0.3cm}
	\parindent=0pt #1
	\end{minipage}}\par}}

\renewenvironment{thebibliography}[1]
	{\begin{list}{\arabic{enumi}.}
	{\usecounter{enumi}\setlength{\parsep}{0pt}
\setlength{\leftmargin 1.25cm}{\rightmargin 0pt}
	 \setlength{\itemsep}{0pt} \settowidth
	{\labelwidth}{#1.}\sloppy}}{\end{list}}

\newcounter{itemlistc}
\newcounter{romanlistc}
\newcounter{alphlistc}
\newcounter{arabiclistc}

\newcommand{\fcaption}[1]{
        \refstepcounter{figure}
        \setbox\@tempboxa = \hbox{\tenrm Fig.~\thefigure. #1}
        \ifdim \wd\@tempboxa > 6in
           {\begin{center}
        \parbox{6in}{\tenrm\baselineskip=12pt Fig.~\thefigure. #1}
            \end{center}}
        \else
             {\begin{center}
             {\tenrm Fig.~\thefigure. #1}
              \end{center}}
        \fi}

\newcommand{\tcaption}[1]{
        \refstepcounter{table}
        \setbox\@tempboxa = \hbox{\tenrm Table~\thetable. #1}
        \ifdim \wd\@tempboxa > 6in
           {\begin{center}
        \parbox{6in}{\tenrm\baselineskip=12pt Table~\thetable. #1}
            \end{center}}
        \else
             {\begin{center}
             {\tenrm Table~\thetable. #1}
              \end{center}}
        \fi}

\def\@citex[#1]#2{\if@filesw\immediate\write\@auxout
	{\string\citation{#2}}\fi
\def\@citea{}\@cite{\@for\@citeb:=#2\do
	{\@citea\def\@citea{,}\@ifundefined
	{b@\@citeb}{{\bf ?}\@warning
	{Citation `\@citeb' on page \thepage \space undefined}}
	{\csname b@\@citeb\endcsname}}}{#1}}

\newif\if@cghi
\def\cite{\@cghitrue\@ifnextchar [{\@tempswatrue
	\@citex}{\@tempswafalse\@citex[]}}
\def\citelow{\@cghifalse\@ifnextchar [{\@tempswatrue
	\@citex}{\@tempswafalse\@citex[]}}
\def\@cite#1#2{{$\null^{#1}$\if@tempswa\typeout
	{IJCGA warning: optional citation argument
	ignored: `#2'} \fi}}

\def\fnt#1#2{\footnotetext{\kern-.3em
	{$^{\mbox{\sevenrm #1}}$}{#2}}}

 1
 1
 1

\font\tenbf=cmbx10
\font\tenrm=cmr10
\font\tenit=cmti10

\font\ninerm=cmr9

\begin{document}
\begin{flushright}
OHSTPY-HEP-T-95-024
\end{flushright}
\vspace{.3in}
\centerline{\tenbf Introduction to Theories of Fermion Masses}
\baselineskip=16pt
\vspace{0.8cm}
\centerline{\tenrm Stuart Raby}
\baselineskip=13pt
\centerline{\tenit Department of Physics, The Ohio State University, 174 W.
18th Ave.}
\baselineskip=12pt
\centerline{\tenit Columbus, OH 43210}
\vspace{0.9cm}
\abstracts{This paper is based on four lectures given at the Trieste Summer
School 1994 on theories of fermion masses.  The first two lectures introduce
three mechanisms which have been used to construct models of fermion masses.
We then discuss some recent applications of these ideas.  In the last lecture
we briefly review SO(10) and some predictive theories of fermion masses.  }

\vfil
\rm\baselineskip=14pt
\section{Introduction}

The Standard Model[SM] provides an excellent description of Nature.
Myriads of experimental tests have to date found no inconsistency.

Eighteen phenomenological parameters in the SM are necessary to fit all
the low energy data[LED]\footnote{This assumes the minimal particle content.
With no right-handed neutrinos and only Higgs doublets, the theory predicts
$m_{\nu} \equiv 0$.}.  These parameters are not equally well known.
$\alpha,
\sin^2(\theta_W),$ $ m_e, m_{\mu}, m_{\tau}$ and $M_Z$ are all   known to
better
than 1\% accuracy.  On the otherhand,  $m_c, m_b,  |V_{us}|$ are known to
between
1\% and 5\% accuracy,  and $   \alpha_s(M_Z), m_u, m_d,$ $ m_s, m _t,
|V_{cb}|,
\left|{V_{ub} \over V_{cb}}\right|,   m_{Higgs}$ and the Jarlskog invariant
measure of CP violation $J$ are   not known to better than 10\% accuracy.   One
of the main goals of the   experimental high energy physics program in the next
5 to 10 years will   be to reduce these uncertainties.  In addition,
theoretical
advances   in heavy quark physics and lattice gauge calculations will reduce
the   theoretical uncertainties inherent in these parameters.  Already the
theoretical uncertainties in the determination of $|V_{cb}|$ from
inclusive B decays are thought to be as low as 5\%\cite{shifman}.
Moreover, lattice calculations are providing additional determinations
of $\alpha_s(M_Z)$ and heavy quark masses\cite{lattice}.

Accurate knowledge of these 18 parameters is important.  They are
clearly not a random set of numbers.  There are distinct patterns which
can, if we are fortunate, guide us towards a fundamental theory which
predicts some (if not all) of these parameters.  Conversely these 18
parameters are the LED which will test any such theory.  Note, that 13
of these parameters are in the fermion sector.  So, if we are to make
progress,  we must necessarily attack the problem of fermion masses.
\newpage
\subsection{Spinor Notation}

There are 15 degrees of freedom in one family of fermions.  We can describe
these states in terms of 15 Weyl spinor fields (each field annihilates a
particle with given quantum numbers and creates the corresponding
anti-particle).  We use the notation   $$ (\nu \;  e) \;\;\; \overline{e}
\;\;\; (u \; d) \;\;\; \overline{u} \;\;\; \overline{d} $$ for
these 15 fields (where the up and down quark fields have an implicit color
index).  The above fields are all left-handed Weyl spinors satisfying the free
field equation of motion (in momentum space) $$ ( E +
\vec{\sigma}\cdot \vec{p} ) \omega(\vec{p}) = 0  $$ where  $ E = |\vec{p}| $
and
$\vec{\sigma}$ are $ 2 \times 2$ Pauli spinors. Rewriting this equation we find
$$ { \vec{\sigma}\cdot \vec{p} \over E } \omega(\vec{p}) = - \omega(\vec{p})
\equiv 2 h \omega(\vec{p}) $$ where   $ h = - 1/2$ is the helicity of the
state.  Thus the states are left-handed ,i.e. their spin is anti-alligned with
their momentum.  We obtain a more compact notation by defining the Lorentz
covariant spinor  $$  \overline{\sigma}_{\mu} \equiv ( 1 , \vec{\sigma} ) .$$
We
then have $ P \kern -8pt / \equiv P^{\mu} \overline{\sigma}_{\mu} = E +
\vec{\sigma}\cdot \vec{p} $.

Note if $\omega(\vec{p})$ is a left-handed spinor, then
$  i \sigma_2 \omega^*(\vec{p})$ satisfying
$${ \vec{\sigma}\cdot \vec{p} \over E } (i \sigma_2 \omega^*(\vec{p})) = +
(i \sigma_2 \omega^*(\vec{p})) $$ is right-handed.

Given the above notation, we can verify that we have accounted
for all the  degrees of freedom in one family.  The field $\nu$ annihilates a
left-handed neutrino and creates a right-handed anti-neutrino, while $\nu^*$
creates a left-handed neutrino and annihilates a right-handed anti-neutrino.
The
CP conjugate of $\nu$ is $\nu_{CP} = i \sigma_2 \nu^*$.

For the electron we need two fields:
e annihilates a left-handed electron and creates a right-handed anti-electron,
while $\oe $ annihilates a left-handed anti-electron and creates a right-handed
electron.  Whereas for the neutrino, only the combined operation CP can be
defined; for the electron we can define the parity operation P such that $e_P =
i \sigma_2 \oe^*$.

It is often useful when calculating Feynman amplitudes to use Dirac notation.
We can always define a Dirac field in terms of two independent Weyl fields.
For the electron we have
$$ \Psi_e = \left(\begin{array}{c} e \\ i \sigma_2 \oe^* \end{array} \right)
.$$
In this basis, the Dirac gamma matrices are given by
$$\gamma_{\mu} = \left( \begin{array}{cc} 0 & \sigma_{\mu} \\
\overline{\sigma}_{\mu} & 0 \end{array} \right) , \; \gamma_5 = \left(
\begin{array}{cc} -1 & 0 \\ 0 & 1 \end{array} \right)$$ with $ \sigma_{\mu} = (
1, -\vec{\sigma} )$.  With this notation the left and right projectors are
given
by $P_{L(R)} = { 1 -(+) \gamma_5 \over 2}$.

Lorentz scalars may be formed in the usual way.  For example, a kinetic term
for
neutrinos is given by  $\overline{\nu}_L \partial \kern -7pt / \nu_L  =  \nu^*
\overline{\sigma}_{\mu} \partial^{\mu} \nu$.  A majorana neutrino mass can be
written as $ \overline{\nu}_L \nu^C_R  + h.c. = \nu \nu + h.c.$ where  $\nu \nu
\equiv \nu i \sigma_2 \nu$.

In the next section we use this formalism when discussing the fermionic sector
of the Standard Model.

\subsection{Standard Model[SM]}

Consider the Yukawa sector of the SM.  We have
$$  {\cal L}_y = {\cal U}^{ij} \ou^0_i h Q^0_j  +  {\cal D}^{ij} \od^0_i \oh
Q^0_j  +  {\cal E}^{ij} \oe^0_i \oh L^0_j  $$
The indices i,j= 1,2,3 label the three fermion families;  ${\cal U}, {\cal D},
{\cal E}$ are complex $3 \times 3$ Yukawa matrices and $h, \oh$ are Higgs
doublets.  In the minimal SM there is only one Higgs doublet and $\oh \equiv i
\tau_2 h^*$.  However in any supersymmetric[SUSY] theory there are necessarily
two independent Higgs doublets, so we will continue to refer to a theory with
two independent Higgs doublets.  The quark and lepton states are defined in
terms of left-handed Weyl spinors and the superscript ``0" refers to the
so-called weak basis in which weak interactions are diagonal.  In Table 1,  we
explicitly define the electroweak charge assignments of all the states.

\begin{table}[t]
\begin{center}
\begin{tabular}{|c|c|c|}
\multicolumn{3}{l}{Table~1. Electroweak Charge Assignments.}\\
 \multicolumn{3}{c}{}\\ \hline\hline
State  & Y -- weak hypercharge & $SU(2)_L$  \\ \hline

$Q^0 = \left( \begin{array}{c} u^0 \\ d^0 \end{array}\right)$ &  1/3 &  doublet
\\
$\ou^0$ & -4/3 & singlet   \\
$\od^0$ & 2/3 &singlet  \\
$L^0 = \left( \begin{array}{c} \nu^0 \\ e^0 \end{array}\right)$ & -1 & doublet
\\
$\oe^0$ & 2 & singlet \\
$ h = \left( \begin{array}{c} h^+ \\ h^0 \end{array}\right) $ & 1 & doublet \\
$ \oh = \left( \begin{array}{c} \oh^0 \\ \oh^- \end{array}\right) $ & -1 &
doublet \\ \hline
\end{tabular}
\end{center}
\end{table}

We also define the vacuum expectation values[vev] of the Higgs fields in the
conventional way $$ \langle h^0 \rangle = {v_u \over \sqrt{2}}, \; \langle
\oh^0
\rangle = {v_d \over \sqrt{2}}$$ with $v = \sqrt{v_u^2 + v_d^2} = 246 GeV$ as
given by the tree relation ${G_F \over \sqrt{2}} = {g_2^2 \over 8 M_W^2} = {1
\over 4 v^2}$ or  $ v = ( 2 \sqrt{2} G_F )^{-1/2}$.  We  also define the
ratio of Higgs vevs $tan\beta \equiv  v_u/v_d$.  Thus fermion masses are given
by
\beqn  m_u \equiv  & {\cal U} sin\beta {v \over \sqrt{2}} & \nonumber \\
        m_d \equiv  & {\cal D} cos\beta {v \over \sqrt{2}} & \nonumber \\
        m_e \equiv  & {\cal E} cos\beta {v \over \sqrt{2}} & \nonumber
        \eeqn

In the weak basis, fermion mass matrices are non-diagonal complex $3 \times 3$
matrices. Note that CP invariance of the SM Lagrangian requires $ m_a = m_a^*$
for a = u, d, e.  Thus a non-removable phase in the fermion Yukawa matrices
violates CP.

 We can always diagonalize the mass matrices with the bi-unitary
transformation $$ m_a^{Diag.} = \overline{V}_a m_a V_a^{\dagger}.$$ The mass
eigenstates are given by $$ u^0 \equiv V_u^{\dagger} u, \;\;\; \ou^0 \equiv \ou
\overline{V}_u $$ and similarly for down quarks and charged leptons.   {\em
There are 9 real (and by convention positive)
mass parameters} given by
$m_u,~m_c,~m_t,~m_d,~m_s,~m_b,~m_e,~m_{\mu},~m_{\tau}$.

In the quark sector, the charged W interactions are given by the term
$ W^{\mu} (u^0)^* \overline{\sigma}_{\mu} d^0 $ (in Dirac notation
$\equiv  W^{\mu} \ou^0_L \gamma_{\mu} d^0_L )$ which in the mass eigenstate
basis becomes $$ W^{\mu}\; u^* V_{CKM} \overline{\sigma}_{\mu} d .$$
The Cabbibo, Kobayashi, Maskawa matrix $V_{CKM}$ is explicitly given by the
expression
\beqn  V_{CKM} & \equiv   ( V_u V_d^{\dagger} ) &
                  =  \left( \begin{array}{ccc}  V_{ud} & V_{us} & V_{ub} \\
                                                 V_{cd} & V_{cs} & V_{cb} \\
                                                 V_{td} & V_{ts} & V_{tb}
                                                 \end{array} \right) \nonumber
                                                 \eeqn
 Note,  $V_{CKM}^{\dagger} V_{CKM} = 1$.
Using the freedom to arbitrarily redefine the phases of all the fermions, the
{\em CKM matrix can be expressed in terms of 4 real parameters (3 real angles
and a CP violating phase).}  {\em There are thus a total of 13 parameters in
the
fermion sector of the SM.}  Note, since neutrinos are massless, we can always
define a basis such that $\nu^0 \equiv V_e^{\dagger} \nu$.  Thus there are no
observable weak mixing matrices in the lepton sector of the theory.

\newpage

\subsection{Summary of Observable Fermionic Parameters}

There is a hierarchy of fermion masses.

$$\begin{array}{ccccc}  \tau (1777 MeV) & > & \mu (105.6 MeV) & > & e (.511
MeV)\\
             b (4.25 \pm .1 GeV) & > & s (150 \pm 30 MeV) & > & d (\sim 7
MeV)\\
             t (174 \pm 17 GeV) & > & c (1.27 \pm 0.05 GeV) & > & u (\sim 5
MeV)
             \end{array} $$

 There is a hierarchy of weak mixing angles as seen in the Wolfenstein
parametrization of the CKM matrix.

 \beqn   V_{CKM} \approx  \left( \begin{array}{ccc}
 1-{\lambda^2 \over 2} & \lambda & A \lambda^3 (\rho + i \eta) \\
 - \lambda & 1-{\lambda^2 \over 2} & A \lambda^2 \\
 A \lambda^3 (1 - \rho + i \eta) & - A \lambda^2 & 1  \end{array} \right)
\nonumber \eeqn

The parameters $\lambda \equiv |V_{us}| \approx .221$ and $A, \rho, \eta \sim
1$.
 Note  $|V_{cb}| = A \lambda^2$ and $\left| {V_{ub} \over V_{cb}} \right| =
\lambda|\rho + i \eta|$.  In this parametrization of the CKM matrix,  $\eta$ is
the CP violating parameter.  However this assignment depends explicitly on the
particular phase convention chosen.  A rephase invariant or convention
independent CP violating parameter is given by the Jarlskog parameter $J$ where
$$  J \equiv  Im (V_{ud}\;V_{ub}^*\;V_{tb}\;V_{td}^*).$$

There is a clear pattern of fermion masses and mixing angles.  We would like to
understand the origin of this pattern.  But no one relation between parameters
can provide that understanding.  It can only come through a quantitative
description of the whole pattern.

\section{Renormalizability and Symmetry}

The 18 phenomenological parameters of the SM are arbitrary independent
renomalized parameters in the SM Lagrangian.  Thus since they are
arbitrary, {\em within the context of the SM they cannot be understood}. They
are merely fit to the data.  The problem of understanding these parameters is
however even worse than you might think. In the fermionic sector of the theory
there are 13 parameters.  Consider however a single charge sector of fermions.
For example, the complex $3 \times 3$ up quark matrix ${\cal U}$ has by itself
18
real arbitrary parameters.   Thus in the fermionic sector there are in
principle many more parameters than there are observables.  This often leads to
much confusion.  In any fundamental theory of fermion masses, we would like to
determine the Yukawa matrices ${\cal U}, \; {\cal D}, \; {\cal E}$.  But only
13
combinations of the 54 parameters in these matrices are observable.  In order
to
understand the pattern of fermion masses, it is necessary to reduce the number
of arbitrary parameters in the Yukawa matrices from 54 to a number which is
less
than 13.

The key ingredients which may allow us to make some progress in this direction
are renomalizable field theories and symmetry.  In a renormalizable field
theory
there are only a finite number of counterterms necessary to define the theory.
For example in QED, we have the renormalized Lagrangian given by $${\cal L} =
Z_2 \overline{\Psi} \partial \kern -7pt / \Psi + Z_1 e \overline{\Psi} A \kern
-7pt / \Psi - Z_3 {1 \over 4} F_{\mu\nu}^2 - Z_m m \overline{\Psi} \Psi .$$  In
this case the electron charge and mass are arbitrary parameters.  However if we
can introduce enough symmetry into a theory such that there are more observable
parameters than there are counterterms, we can in principle obtain predictable
relations among these parameters.

In the following, we will discuss three mechanisms which have been used in the
past for obtaining relations between fermion masses and mixing angles.  We will
then discuss more recent realizations using these 3 tools of the trade.

\subsection{Tools of the Trade}

Before describing each mechanism in detail, let me give a brief description of
the seminal ideas involved.  We will broadly classify the 3 mechanisms
as {\em radiative}, {\em textures} and {\em effective operator} relations.

\begin{itemize}
\item {\em Radiative}  In this example,  we calculate the electron mass as a
radiative correction proportional to the muon mass. We show that the gauge
symmetry of the theory allows only one Yukawa coupling for both $\mu$ and $e$.
In addition, as a consequence of a missing vacuum expectation value[vev], the
muon obtains mass at tree level while the electron remains massless.  At one
loop we then find  $m_e \sim  \alpha m_{\mu}$.

\item {\em Textures}  We use both gauge and discrete family
symmetries to define the most general Yukawa matrix for a pair of quarks which
is symmetric and has a certain number of zero elements,  thereby reducing the
number of fundamental parameters.  We thus obtain tree level relations among
quark masses and mixing angles.  Note since experimentally  $m_d/m_s \sim 1/20
>> \alpha$, it would not be possible to obtain all mass ratios radiatively.

\item {\em Effective Operators}  We use U(1) symmetries with light fermions
coupled to heavy fermions with mirror partners.  When integrating the heavy
fermions out of the theory we generate effective higher dimension fermion mass
operators which explain the fermion mass hierarchy.

\end{itemize}

\subsection{Radiative mechanism}[Weinberg, 1972;  Georgi and Glashow,
1973]\cite{weinberg}  In a seminal paper, coming shortly after the proof of the
renormalizability of spontaneously broken non-abelian gauge
theories\cite{thooft}, Weinberg emphasized the advantages of renormalizable
field theories for obtaining calculable fermion masses.  In a simple example he
showed that the electron mass can be generated radiatively from the muon mass.
There was a critical flaw in his example which was later pointed out and
corrected by Georgi and Glashow.

Consider a theory describing just two families of leptons.  The electroweak
gauge group is extended to $G_W = SU(3)_1 \times SU(3)_2$ which allows only one
Yukawa coupling $\lambda$ for both $\mu$ and $e$.  In addition the theory has a
discrete parity invariance which interchanges the states transforming under the
two SU(3)s, i.e. $1 \leftrightarrow 2$ which among other things allows only
one gauge coupling constant, $g$. The fermions are represented by  $$ \Psi_1
\equiv \left(\begin{array}{c} \overline{\mu} \\ \nu_e \\ e \end{array}\right) ,
\;\;  \Psi_2 \equiv \left(\begin{array}{c} \overline{e} \\ \nu_{\mu} \\ \mu
\end{array}\right)$$  which are in the (3,1), (1,3) representation
of $G_W$, respectively. The minimal Higgs content $\Phi_{ab} =
(\overline{3},\overline{3}), \; a,b = 1,2,3$ contains two Higgs doublets when
looked at in terms of the $SU(2)_L \times U(1)_Y$ subgroup of $G_W$.  The
$SU(2)_L \times U(1)_Y$ subgroup is explicitly defined by the generators $T_i =
t_i^1 + t_i^2, \; i=1,2,3$ and $ Y = - \sqrt{12} (t_8^1 + t_8^2)$ with $$t_8
\equiv {1 \over \sqrt{12}} \left( \begin{array}{ccc} -2 & 0 & 0\\ 0 & 1 & 0\\ 0
& 0 & 1 \end{array}\right).$$ The only renormalizable Yukawa coupling is given
by
$$\lambda \Psi_1^a \Phi_{ab} \Psi_2^b.$$  As a result, the $\mu, e$ masses are
given in terms of the expressions $m_{\mu} = \lambda v_{\mu}, \; m_e = \lambda
v_e$ where $v_{\mu} = \langle \Phi_{13}\rangle, \; v_e = \langle \Phi_{31}
\rangle$ are the two vevs of $\Phi$ which break $SU(2)_L \times U(1)$ to
$U(1)_{EM}$.  The most general renormalizable potential for $\Phi$ is defined
such that, for a finite range of parameters, the minimum energy state has
$v_{\mu} \neq 0, v_e = 0$.  Thus the electron is massless at tree level.
However there is no symmetry which can protect  the electron from obtaining a
mass radiatively since the chiral symmetries of $e$ and $\mu$ are united by the
gauge group $G_W$.  In fig. 1 we show the Feynman diagram which contributes to
the electron mass.

The problem with this model, discovered by Georgi and Glashow, is evident from
the Feynman diagram of fig. 2.  This diagram is obtained by closing the
external
fermion line in fig. 1.  This diagram is logarithmically divergent.  It in fact
generates the local dimension 4 operator  $$\Phi_{31}^* \Phi_{13} X X^* .$$
Such
a term must be in the Lagrangian since there is clearly no symmetry which
prevents it and it is a dimension 4 operator which requires a fundamental
parameter to renormalize.  This term has the nasty effect of driving $v_e =
\langle \Phi_{31} \rangle \neq 0$.  In order to solve this problem and have a
renomalizable scalar potential such that $v_e = 0$ ``naturally",  Georgi and
Glashow proposed to enlarge the gauge symmetry $G_W$ further.  The details are
not important.  It is important to recognize however that the problem for
Weinberg's example is that the most general renormalizable potential for $\Phi$
did not satisfy the requirement that, for a finite range of parameters, the
minimum energy state has $v_e = 0$.

\subsection{Textures} [Weinberg; Wilczek and Zee; Fritzsch,
1977]\cite{fritzsch}
In 1968, several people made the observation of a simple empirical relation
between the Cabibbo angle and the down and strange quark mass ratio given
by\cite{gatto} $$ tan\Theta_c = \sqrt{{m_d \over m_s}} \approx {f_{\pi} m_{\pi}
\over f_K m_K}.$$  It was not until 9 years later that a possible explanation
of
this relation was proposed.\cite{fritzsch}

Before we discuss the explanation, let's consider the general problem.  The up
and down quark mass terms, defined in the weak eigenstate basis, are given by
$$\delta {\cal L} = (\begin{array}{cc} \ou & \oc \end{array}) m_u \left(
\begin{array}{c} u \\ c \end{array} \right)  + (\begin{array}{cc} \od & \os
\end{array}) m_d \left( \begin{array}{c} d \\ s \end{array} \right)$$ where in
general the up and down mass matrices are given by $$m_u = \left( \bar{cc}
\tilde{C} & \tilde{B'} \\ \tilde{B} & \tilde{A} \ear \right), $$  $$m_d =
\left(
\bar{cc} C & B' \\ B & A \ear \right).$$  $A, B, B', C,
\tilde{A},\tilde{B},\tilde{B'},\tilde{C}$ are in general arbitrary complex
parameters.  Note however that not all the phases are physical.  We can
redefine
the phases of the fields $u, \ou,$ $c, \oc,$ $ d, \od,$ $ s, \os$ and remove 5
of
these phases {\em without introducing any new phases anywhere else in the
Lagrangian}. For example, redefine the phase of $\os$ to make $B$ real, then
redefine the phase of $s$ to make $A$ real.  Note that we must also redefine
the
phase of $c$ by the same amount as $s$ so that we don't introduce a new phase
in
the W-c-s vertex.  Next redefine the phases of $\od, \oc, \ou$ making $B'$ and
$\tilde{A}$ real and $arg \tilde{B} = - arg \tilde{B'}$.  We now see
that the mass eigenstates and mixing angles in the up (down) sector depend on 6
(5) parameters for a total of 11 parameters.  However, how many observables are
there?  There are 4 quark masses and one electroweak mixing angle or a total of
5 parameters.  We certainly have enough arbitrary parameters to fit these 5
observables, but we are not able to make any predictions.  In order to make
predictions we must reduce the number of arbitrary parameters.   In order to
reduce the number of fundamental parameters we need to introduce symmetries.
In
the  paper by Fritzsch (see (\cite{fritzsch})) it was shown that by
\begin{itemize}
\item extending the electroweak gauge symmetry to $SU(2)_L \otimes SU(2)_R
\otimes U(1)$,  and
\item demanding Parity, CP and an additional discrete
symmetry
\end{itemize} the number of arbitrary parameters in $m_u, m_d$ can be
reduced to 4.  This allows for one prediction which relates masses and the one
mixing angle.  The discrete symmetry enforces $C = \tilde{C} = 0$ and parity
requires the matrices be Hermitian.  The resulting matrices have the form $$m_u
= \left( \bar{cc} 0 & |\tilde{B}| \\ |\tilde{B}| & |\tilde{A}| \ear \right), $$
$$m_d = \left( \bar{cc} 0 & |B|e^{-i\gamma} \\ |B|e^{i\gamma} & |A| \ear
\right).$$  Note without assuming CP (which requires $m_u$ and $m_d$ to be
real)
we remain with one phase and lose the prediction.

Let me now give Fritzsch's model in more detail.  The model included 2
left-handed quark and 2 left-handed anti-quark doublets $$ Q_i = \left( \bar{c}
u \\ d \ear \right)_i,  \;\; \overline{Q}_i = \left( \bar{c} \ou \\ \od \ear
\right)_i, \; i=1,2$$ transforming in the $(2,1), (1,\overline{2})$
representations of $SU(2)_L \otimes SU(2)_R \otimes U(1)$ with equal and
opposite U(1) charges.  The index i is a generation label.  Thus you should
consider the quarks denoted by 2 as c and s quarks and those by 1 as u and d.
In addition, the model has 2 scalar multiplets $\phi_{1(2)}$ in the
$(\overline{2},\overline{2})$ representation.  Without any additional
symmetries
the allowed scalar -quark -anti-quark couplings are given by $$\delta {\cal L}
=
\lambda_{ij} \overline{Q}_i \phi_1 Q_j + \lambda'_{ij} \overline{Q}_i \phi_2
Q_j
$$  with $\lambda_{ij},\lambda'_{ij}$ arbitrary complex coupling constants.  If
we now imposed CP invariance on the Lagrangian, then $\lambda_{ij}$ and
$\lambda'_{ij}$ are real.  Under Parity  $$ Q \leftrightarrow i \sigma_2
\overline{Q}^*,  \; \phi \leftrightarrow  \sigma_2 \phi^{\dagger} \sigma_2 .$$
Imposing P on the Lagrangian requires $$\lambda_{ij} = \lambda^*_{ji}, \;
\lambda'_{ij} = (\lambda'_{ji})^*.$$  Finally, we define two additional
discrete
symmetries $\{ P_1, P_2 \}$ which act on the set of fields in the following
way.
$$ P_1 : \bar{c} \{ Q_1, \overline{Q}_1, \phi_1 \} \rightarrow  (-1)\times \{
Q_1, \overline{Q}_1, \phi_1 \}\\ \{ Q_2, \overline{Q}_2, \phi_2 \} \rightarrow
(i)\times \{ Q_2, \overline{Q}_2, \phi_2 \} \ear$$ $$ P_2 :  \{ Q_2,
\overline{Q}_2, \phi_2 \} \rightarrow  (-1)\times \{ Q_2, \overline{Q}_2,
\phi_2
\}.$$

I list below the only terms in $\delta {\cal L}$ allowed by $P_2$ --
$$ \lambda'_{12} ( \overline{Q}_2 \phi_2 Q_1 + \overline{Q}_1 \phi_2 Q_2 )
+ \lambda_{22} (\overline{Q}_2 \phi_1 Q_2) + \lambda_{11} (\overline{Q}_1
\phi_1
Q_1).$$
If we now impose $P_1$ we are left with
$$\lambda'_{12} ( \overline{Q}_2 \phi_2 Q_1 + \overline{Q}_1 \phi_2 Q_2 )
+ \lambda_{22} (\overline{Q}_2 \phi_1 Q_2).$$  The up and down quark mass
matrices are now
given by
$$ m_u = \left( \bar{cc} 0 & \lambda'_{12} \langle \phi_2^u \rangle \\
                   \lambda'_{12} \langle \phi_2^u \rangle &   \lambda_{22}
\langle \phi_1^u \rangle  \ear \right) \equiv \left( \bar{cc} 0 & \tilde{B}\\
                   \tilde{B} & \tilde{A} \ear \right) $$
and $m_d$ is given by the same expression with $\phi_{1(2)}^d$ replacing
$\phi_{1(2)}^u$
or
$$ m_d \equiv \left( \bar{cc} 0 & B\\ B & A \ear \right).$$
Note $\phi^{u(d)}$ are the neutral components of the scalar $\phi$ which give
mass to up (down) quarks.  The weak vev $v$ is given by $ v =
\sqrt{(\phi_1^u)^2
+ (\phi_2^u)^2 +(\phi_1^d)^2 +(\phi_2^d)^2 }$.

We can now obtain the successful relation
$$ tan\Theta_c \approx ( \sqrt{{m_u \over m_c}} - \sqrt{{m_d \over m_s}} ). $$

\subsection{Effective Operators}[Froggatt and Nielsen, 1979]\cite{froggatt}
In the previous mechanism, the small mass ratios $m_u/m_c (m_d/m_s)$ are given
in terms of
arbitrary ratios $ \tilde{B}^2/\tilde{A}^2 (B^2/A^2)$.  But we have no
understanding of
why $\tilde{B} << \tilde{A}$, etc.   Froggatt and Nielsen tried to provide this
explanation.

Consider the SM with an additional global U(1) symmetry denoted by Q.   The
quantum numbers of , for example, up-type quarks under Q are given by
$\overline{q}_i \equiv q(\ou_i),\;\; q_i \equiv q(u_i)$ and we take $q(Higgs)
\equiv 0$.  We assume that Q is spontaneously broken and that the symmetry
breaking is communicated to quarks by the insertion of a tadpole with magnitude
$\epsilon < 1$ and charge -1.  It is then assumed that $\overline{q}_3 = q_3 =
0$ with $\overline{q}_i,\;\; q_i$ non-vanishing such that the Yukawa term
$\ou_3
h u_3$ is the only Q invariant term without a symmetry breaking insertion.  The
term $\ou_i h u_j$ has Q charge $\overline{q}_i+q_j$ and needs the insertion
$\epsilon^{\overline{q}_i+q_j}$ to be invariant (see fig. 3).  These
effective higher dimension operator terms are thus suppressed with respect to
the direct dimension four Yukawa coupling.

What is the origin of the small parameter $\epsilon$?  Consider the graphs of
fig. 4 which describes a simple two family quark model.  We have introduced two
new scalars $\phi^0, \phi^{-1}$, singlets under the electroweak symmetry with
Q
charge denoted by the superscript and the  left-right symmetric up-type quarks
$U^{\pm 1}, \overline{U}^{\pm 1}$, members of an $SU(2)_L$ doublet,
anti-doublet, respectively.  Both $\phi^0, \phi^{-1}$ are  assumed to get
non-vanishing vevs satisfying $\phi^0 > \phi^{-1} >> M_Z$.  As a result, the
new
up-type quarks are heavy with mass of order $\langle \phi^0 \rangle$.  Due to
the
expectation values of the weak Higgs $h^0$ and the new scalar $\phi^{-1}$,
these
heavy quarks mix with the light quarks.  Fig. 4  represents this mixing.  These
graphs generate off-diagonal mixing in the fermion mass matrices between the
light up quarks of the second and third family. We have  $\epsilon (\ou^1_2 h
u^0_3 + \ou^0_3 h u^1_2)$ as the lowest order mixing obtained in a power series
expansion in the small parameter $\epsilon = {\langle \phi^{-1} \rangle \over
\langle \phi^0 \rangle}$.

The procedure of reading the low energy mixing term off of the diagram of fig.
4  is equivalent to the procedure of diagonalizing the fermion mass matrix,
ignoring the weak vev of $h$.  For example, consider the mass terms represented
as vertices in fig. 4   for the state $\overline{U}^{+1}$.  We have $$
\overline{U}^{+1} ( \langle \phi^0 \rangle U^{-1} + \langle \phi^{-1} \rangle
u^0_3 ) \approx \langle \phi^0 \rangle \overline{U}^{+1} ( U^{-1} + \epsilon
u^0_3 ).$$  Define the massive state  $u_M \approx  U^{-1} + \epsilon u^0_3 $
and the orthogonal massless state is  $u'^0_3 \approx - \epsilon U^{-1} +
u^0_3$.  At energies much below $\langle \phi^0 \rangle$ and greater than $
\langle h \rangle $ we can define an effective theory by integrating out the
states with mass of order $\langle \phi^0 \rangle$.  In this effective theory,
the vertex $\ou_2^1 h U^{-1}$ becomes $ - \epsilon \ou_2^1 h u'^0_3$ which is
obtained by using the relation $U^{-1} \approx - \epsilon u'^0_3 + u_M$.  Of
course, the exact effective dimension 4 Yukawa coupling (which contains an
expansion in $\epsilon$) is obtained by using the exact expressions for the
massive and massless eigenstates.

In this mechanism the extra global symmetry Q controls the textures of
effective mass operators in fig. 3.

\section{Theories of Fermion Masses - Survey (1979 - 1994)}

In the last 15 years, there have been many papers on fermion masses.  Most of
these papers, if not all of them, have been applications of one or more of the
mechanisms or tools for fermion masses we discussed in the previous section.
In
this section I will consider a few representative examples of papers in the
literature.  I make no claim that these examples are all inclusive.

\subsection{Radiative mechanism}
The extended technicolor theory of fermion masses assumes that the light quarks
and leptons receive their mass via a radiative mechanism from new heavy
technifermions.  The technifermion mass, on the otherhand, results from a
chiral symmetry breaking condensate due to new strong technicolor
interactions[see Dimopoulos and Susskind, Eichten and Lane]\cite{technicolor}.
These models are notoriously non predictive as a result of the strong
interactions which are needed for chiral symmetry breaking.  One can at best
obtain order of magnitude estimates for quark masses given by formulae such as
$m_q \sim  {\langle \overline{T} T \rangle \over \Lambda^2_{ETC}}$ where
$\langle  \overline{T} T \rangle$ is the technifermion condensate and
$\Lambda_{ETC}$ is the ETC breaking scale.

In recent models, people have attempted to get fermion masses in SUSY theories
by feeding masses from the squark and slepton sector into the quark and lepton
sectors\cite{susyradiative}.  For example, in the graph of fig. 5 the down
quark gets mass from a soft SUSY breaking bottom squark mass squared given
by $A m_b$.   This leads to a down quark mass  $$m_d \approx {\alpha_s \over 2
\pi} \left( {\delta \tilde{m}^2 \over \tilde{m}^2}\right)^2\left( {A
m_{\tilde{g}} \over \tilde{m}^2}\right) m_b$$ where $\delta \tilde{m}^2$ is a
measure of the bottom and down squark mixing in the quark-squark basis which
diagonalizes quark masses at tree level.  Since such a theory replaces the
arbitrary Yukawa parameters in the fermion sector by new arbitrary mass
parameters in the scalar sector, it is not clear that one can really make
progress using this paradigm.

\subsection{Textures and Discrete Symmetries}

Fritzsch generalized his theory of the Cabibbo angle to a complete 6 quark
model\cite{fritzsch2}.  This model has 6 real magnitudes and 2 phases or eight
parameters to fit 10 observables (six quark masses and 4 CKM angles).  There
are
thus 2 predictions.  One of these predictions, as shown by Gilman and
Nir\cite{fritzsch2}, is that the top quark is necessarily light, i.e. $m_t < 96
GeV$.  Hence the Fritzsch texture is now ruled out by experiments at Fermilab.

\subsection{Effective Operators}

A SUSY version of the Froggatt and Nielsen mechanism has recently been studied
in the literature\cite{leurer} within the context of the SM gauge group.  In
these models the fermion mass matrices have the form $$m_{ij} =
\epsilon^{q(\overline{f}_i) + q(f_j)}$$ where  $f = u, d, e$.  This paradigm
can
``naturally" explain the zeros in mass matrices and certain order of magnitude
ratios of non-vanishing elements, but unfortunately it has no power to predict
testable fermion mass relations.  The proof of this paradigm would be found in
the existence of new states with mass above the weak scale responsible for the
effective operators.

All of the examples discussed so far have the following features in common:
\begin{enumerate}
\item They are all relations defined just above the weak scale; as a
consequence
they  all require new physics (new particles and/or gauge symmetries) just
above
experimental observation. \item They all (except for the Fritzsch Ansatz) give
only a qualitative description of fermion masses;  thus there are no testable
predictions for fermion masses and mixing angles. \item Quark and lepton masses
are unrelated. \end{enumerate}

An important question is what is the scale of new physics; the scale at which
new
symmetries and particles appear.  If this new scale is just
above the weak scale then we must worry about possible new flavor changing
neutral current[FCNC] interactions.  In radiative mechanisms,  loop diagrams
can
contribute to new FCNC interactions (for example see fig. 6).  In this case the
effective FCNC interactions are of order $$\delta {\cal L} \sim \alpha_W^2
\left({\delta \tilde{m}^2 \over \tilde{m}^2}\right) {1 \over \tilde{m}^2}
(s^*d)^2. $$  They are proportional to squark mixing mass terms and can be
suppressed by increasing the overall squark mass scale.  In the texture
mechanism, the new states required to incorporate the necessary discrete and
gauge symmetries which make texture zeros ``natural" will contribute to FCNC
interactions.  Finally in the Froggatt-Nielsen mechanism, the new heavy
fermions
and scalars can also contribute to FCNC interactions.  In all cases, one must
compare the new FCNC interactions with experiment and place bounds on the scale
of new physics.   Generically, these bounds will force the new physics scale to
be in the $(1 - 10^3) TeV$ range depending in detail on the specific process
considered.

\section{SUSY GUTs}

We would like to obtain models {\em more predictive} than our previous
examples.
In order to do this we need {\em more symmetry}.  We can gain a lot of
predictive power by relating quark and lepton masses.  Of course this requires
some sort of grand unification symmetry\cite{gg}.

In the rest of these lectures I will consider the consequences of SUSY Grand
Unified Theories [GUTs]\cite{gqw,drw1}.  The main reason is that they already
make one prediction which agrees remarkably well with low energy
data\cite{recent}.   Using the measured values of $\alpha$ and $sin^2\theta_W$,
and assuming reasonable threshold corrections at the weak and GUT scales,
Langacker and Polonsky\cite{langacker} obtain the prediction for
$\alpha_s(M_Z)$
in fig. 7.    They also plot the experimental measurements of $\alpha_s(M_Z)$
and
you can see that the two are in remarkable agreement.  Note that the minimal
non-SUSY GUT gives a value for $\alpha_s(M_Z) \sim 0.07$ which is several
standard deviations away from the observations.

Let us now consider the first predictions from GUTs for fermion masses.  In
order
to do this we will give a brief introduction to $SU(5)$\cite{gg}.   The quarks
and leptons in one family of fermions fit into two irreducible representations
of $SU(5)$: $10_{ij} = - 10_{ji}, \;\; \overline{5}^i = \epsilon^{ijklm}
\overline{5}_{jklm}$ where $i,j,k,l,m = 1 - 5$ are $SU(5)$ indices.  In the
fundamental 5 dimensional representation of $SU(5)$ the adjoint is represented
by $5 \times 5$ traceless hermitian matrices. We can consider the indices from
$1 - 3$ as being color indices acted on by the $SU(3)_{color}$ subgroup of
$SU(5)$ and the indices $4,5$ as weak $SU(2)_L$ indices.  Hypercharge is
represented by the matrix $$ Y = - 2\sqrt{{5 \over 3}} Y_5 ~{\rm and}~ Y_5
\equiv {1 \over \sqrt{60}} \left( \bar{ccccc} 2 & 0 & 0 & 0 & 0 \\ 0 & 2 & 0 &
0
& 0 \\ 0 & 0 & 2 & 0 & 0 \\ 0 & 0 & 0 & -3 & 0 \\ 0 & 0 & 0 & 0 & -3 \ear
\right)$$ satisfying $Tr Y_5^2 = 1/2$. From this embedding of the SM into
$SU(5)$ we can check that the states fit into the $10$ and $\overline{5}$ as
follows: $$ 10 =  \left( \bar{cc}  \ou & Q \\  & \oe \ear \right), \;\;
\overline{5} = \left( \bar{c} \od \\ L \ear \right).$$ The two Higgs doublets
fit into a $5 (\equiv H)$ and $\overline{5} (\equiv \overline{H})$.  Similarly
$H$ and $\overline{H}$ can be decomposed into weak doublets and color triplets
under the SM symmetry.   We have $$\overline{H}  = \left( \bar{c} \ot \\
\overline{h} \ear \right), \;\; H = \left( \bar{c} t \\ h \ear \right)$$ with
$t
(h)$ denoting triplet(doublet) states.

Up and down quark Yukawa couplings at $M_{GUT}$ are given in terms of the
operators  $$\lambda_u H_i 10_{jk} 10_{lm} \epsilon^{ijklm} + \lambda_d
\overline{H}^i 10_{ij} \overline{5}^j.$$  When written in terms of quark and
lepton states we obtain the Yukawa couplings to the Higgs doublets $$\lambda_u
\ou h Q + \lambda_d ( \od \overline{h} Q + \oe \overline{h} L).$$  We see that
$SU(5)$ relates the Yukawa couplings of down quarks and charged leptons, i.e.
$\lambda_d = \lambda_e$ at the GUT scale.  \underline{Assuming this relation
holds for all 3 families}, we have\cite{chanowitz} $\lambda_b =
\lambda_{\tau},\;\; \lambda_s = \lambda_{\mu},\;\; \lambda_d = \lambda_e$ at
$M_{GUT}$.

To compare with experiment we must use the renormalization group[RG] equations
to run these relations (valid at $M_{GUT}$) to the weak scale.  The first
relation gives a prediction for the b-$\tau$ ratio which is in good
agreement with low energy data. Note, for heavy top quarks we must now use the
analysis which includes the third generation Yukawa couplings\cite{inoue}.  We
will discuss these results shortly.  The next two relations can be used to
derive the relation: ${\lambda_s \over \lambda_d} = {\lambda_{\mu}
\over \lambda_e}$ at $M_{GUT}$.  However at one loop the two ratios are to a
good approximation RG invariants.  Thus the relation is valid at any scale $\mu
< M_{GUT}$.  This leads to the {\em bad} prediction  $${m_s \over m_d} =
{m_{\mu} \over m_e}$$ for running masses evaluated at 1 GeV.  It is a bad
prediction since experimentally the left hand side is $\sim 20$ while the rhs
is
$\sim 200$.

An ingenious method to fix this bad relation was proposed by Georgi and
Jarlskog\cite{gj}.  They  show how to use $SU(5)$ Clebschs in a novel texture
for fermion Yukawa matrices to keep the good relation $\lambda_b =
\lambda_{\tau}$, and replace the bad relation above by the good relation $${m_s
\over m_d} = {1 \over 9} {m_{\mu} \over m_e}.$$

\subsection{Georgi-Jarlskog Texture}

Georgi and Jarlskog found an interesting texture which resolved the problem of
light fermion masses.  They also constructed a grand unified theory with 3
families of quarks and leptons, the necessary Higgs and with sufficient
symmetry
so that the theory was ``natural."

The fermion Yukawa matrices have the form
$$ {\cal U} =  \left( \bar{ccc} 0 & C & 0 \\ C & 0 & B \\ 0 & B & A \ear
\right), \;\;
{\cal D} =  \left( \bar{ccc} 0 & F & 0 \\ F' & E & 0 \\ 0 & 0 & D \ear \right),
\;\; {\cal E} =  \left( \bar{ccc} 0 & F' & 0 \\ F & -3 E & 0 \\ 0 & 0 & D \ear
\right) $$ where $ A, B, C, D, E, F, F'$ are in general arbitrary complex
parameters.  The $SU(5)$ version of the theory contains in addition to the
Higgs
multiplets, $ \overline{H} = \overline{5}, \; H = 5$ discussed previously, a
$\overline{45}$.  The Yukawa Lagrangian is given by $$\overline{H} ( F' 10_1
\overline{5}_2 + F 10_2 \overline{5}_1 ) + D \overline{H} 10_3 \overline{5}_3
+
E \overline{45} 10_2 \overline{5}_2 $$ $$ + C H 10_1 10_2 + B H 10_2 10_3 + A H
10_3 10_3 .$$   Note if we diagonalize the down and charged lepton matrices in
the $2 \times 2$ subspace of the two light generations we find the relations $
\lambda_s \approx {1 \over 3} \lambda_{\mu}, \;\;  \lambda_d \approx 3
\lambda_e$ resulting from the Clebsch factor of 3.  This factor of 3 is very
natural in any GUT since it just results from the fact that there are three
quark states for every lepton state.  After RG running from $M_{GUT}$ to 1 GeV
we obtain the good mass relations $ m_s \approx {4 \over 3} m_{\mu}, \;  m_d
\approx 12 m_e$.

Note, the up mass matrix is necessarily symmetric but within $SU(5)$ the down
matrix is not.   It was shown by Georgi and Nanopoulos\cite{gjso10} that by
extending the gauge symmetry to $SO(10)$ the down matrix will also be
symmetric.  In this case a 126 dimensional representation is needed to obtain
the
Clebsch of 3.  A complete SO(10) version of the theory was first given in a
paper by Harvey, Ramond and Reiss\cite{gjso10}.

Since SUSY GUTs seem to work so nicely for gauge coupling unification, it is
natural to wonder whether a SUSY version of the Georgi-Jarlskog ansatz gives
reasonable predictions for fermion masses and mixing angles.  Dimopoulos, Hall
and I showed that the predictions for fermion masses and mixing angles worked
very well\cite{dhr,dhr90}.  Using the freedom to redefine the phases of
fermions, we showed that there were just 7 arbitrary parameters in the Yukawa
matrices; a standard form is given by $$ {\cal U} =  \left( \bar{ccc} 0 & C & 0
\\ C & 0 & B \\ 0 & B & A \ear \right), \;\;  {\cal D} =  \left( \bar{ccc} 0 &
F
e^{i \phi} & 0 \\ F e^{-i \phi} & E & 0 \\ 0 & 0 & D \ear \right), \;\; {\cal
E}
=  \left( \bar{ccc} 0 & F e^{i \phi} & 0 \\ F e^{-i \phi}& -3 E & 0 \\ 0 & 0 &
D
\ear \right) $$ where now $A, B, C, D, E, F, \phi$ are the 7 real parameters.
Including $tan\beta$ we have 8 real parameters in the fermion mass matrices.
On the otherhand, there are 14 low energy observables,  9 charged fermion
masses,
4 quark mixing angles and $tan\beta$;  thus there are 6 predictions.

We used the best known low energy observables, $m_e, m_{\mu}, m_{\tau}, m_c,
m_b, |V_{us}|, {m_u \over m_d}$ as input to make predictions for $m_t, m_s,
|V_{cb}|, m_d, \left| {V_{ub} \over V_{cb}} \right|$ and the CP violating
parameter $J$ in terms of arbitrary values of $tan\beta$.  The results were in
good agreement with the low energy data.   Fitting all the parameters
simultaneously, Barger, Berger, Han and Zralek\cite{dhr90} showed this texture
agreed with all the low energy data at 90\% CL.  Recently Babu and Mohapatra
have found an interesting representation for the Georgi-Jarlskog texture in
terms of an effective theory at $M_{Planck}$\cite{babumohapatra}.  This is an
SO(10) theory containing effective mass operators with dimension $\ge 4$ which
eliminates the need for the large 126 dimensional representation.

To conclude this review of the literature,  other textures have recently been
pursued.  Different SO(10) SUSY GUT textures have been
discussed\cite{so10ansatze}.  Babu and Shafi have considered the SUSY version
of
Fritzsch (defined at a GUT scale) and showed that $m_t < 120
GeV$\cite{babushafi}.  Finally Ramond, Roberts and Ross have, in a bottom-up
approach, classified all symmetric quark mass matrices within the minimal
supersymmetric standard model[MSSM] with texture zeros at $M_{GUT}$\cite{rrr}.
They find 6 solutions which fit the data.

\subsection{Renormalization Group Running}

In this section I want to discuss the RG equations for the bottom, top and tau
Yukawa couplings in SUSY GUTs.  The RG equations described below neglect mixing
among the different generations.  For the light families one can neglect the
effect of the Yukawa couplings in the beta functions on the right-hand-side of
these equations.  Specifying a typical Yukawa coupling by $\lambda$ we define
the quantity $Y \equiv \lambda^2/(4\pi)^2$.  We also define $\tilde{\alpha}_i
\equiv \alpha_i/(4 \pi)$ and $ t = ln {M_G^2 \over \mu^2}$.  In terms of these
parameters the RG equations are\cite{inoue}:

$${d Y_t \over dt} = Y_t ( {16 \over 3} \tilde{\alpha}_3 + 3 \tilde{\alpha}_2 +
{13 \over 9} \tilde{\alpha}_1 - 6 Y_t - Y_b - Y_{\nu_{\tau}} \theta ) ,$$
$${d Y_b \over dt} = Y_b ( {16 \over 3} \tilde{\alpha}_3 + 3 \tilde{\alpha}_2 +
{7 \over 9} \tilde{\alpha}_1 -  Y_t - 6 Y_b - Y_{\tau} ) ,$$
$${d Y_{\tau} \over dt} = Y_{\tau} ( 3 \tilde{\alpha}_2 + 3 \tilde{\alpha}_1
-3 Y_b - 4 Y_{\tau} - Y_{\nu_{\tau}} \theta) ,$$ and
$${d Y_{\nu_{\tau}} \over dt} = Y_{\nu_{\tau}} (  3 \tilde{\alpha}_2 +
{3 \over 5} \tilde{\alpha}_1 - 3 Y_t - 4 Y_{\nu_{\tau}} \theta  - Y_{\tau} ).$$

We have included the RG equations for the tau neutrino assuming a Dirac mass
term for the left handed tau neutrino coupled to a singlet state.  The function
$\theta$ is zero for $t > ln {M_G^2 \over M^2}$ and one otherwise, where $M$ is
the majorana mass
of the singlet neutrino.  For $M \sim M_{GUT}$ the tau neutrino does not affect
the running of the charged fermions.

For light quarks and leptons it is easy to see that the additional color
interactions for quarks explains why the ratio of quark to lepton Yukawa
couplings increases at low energy.  For the bottom to tau mass ratio this
increase is in fact too large if one begins with the unification assumption
that
$\lambda_b = \lambda_{\tau}$ at $M_{GUT}$.  It was shown by Inoue et. al. and
Iba\~{n}ez and Lopez that a large top quark Yukawa coupling can decrease the
ratio $\lambda_b/\lambda_{\tau}$ at low energies\cite{inoue}\footnote{In a
one Higgs model, however, the top quark Yukawa coupling tends to increase the
ratio $\lambda_b/\lambda_{\tau}$}.     In fig. 8  we show this ratio as a
function of the top quark mass valid for small $tan\beta$ or equivalently
neglecting $\lambda_b$ and $\lambda_{\tau}$ in the RG runnning.  In fig. 9 we
show the prediction for the top quark running mass as a function of
$tan\beta$ assuming $b - {\tau}$ unification at $M_{GUT}$\cite{recent}.  You
see that the top
quark is naturally heavy and can easily be in the range observed at Fermilab.

As an aside, it has been noted recently by several authors\cite{vissani} that
the
tau  neutrino can affect the RG equations significantly if its mass is in the
few eV range making it a good candidate for a hot component to the dark matter
in the universe.  In this case $M \sim 10^{12} GeV$ and the tau neutrino
becomes
important.  They noticed that the tau neutrino offsets the effect of the top
quark Yukawa coupling to decrease the bottom to tau mass ratio.  See for
example
the equation below.

$${d  \over dt}\left( {Y_b \over Y_{\tau}} \right) = \left( {Y_b \over
Y_{\tau}}
 \right) ( {16 \over 3} \tilde{\alpha}_3 - {20 \over 9} \tilde{\alpha}_1 -
(Y_t
- Y_{\nu_{\tau}} \theta) - 3 ( Y_b - Y_{\tau} )) .$$

In order to affect a significant decrease they show that the bottom quark
Yukawa
coupling, which also tends to drive the bottom to tau ratio down, must be
significant, requiring values of $tan\beta$ larger than about 10.

Finally Ananthanarayan, Lazarides and Shafi\cite{als} have studied the SO(10)
GUT
boundary conditions $\lambda_t = \lambda_b = \lambda_{\tau} $.  They have
demonstrated that these conditions are consistent with the low energy data.
They necessarily require large values of $tan\beta \sim 50$.   We will study
this case in more detail, but first let me briefly discuss the group SO(10).

\section{Introduction to SO(10) Group Theory}

The defining representation is a ten dimensional vector denoted by
$ 10_i ,  \;\; i = 1 , \cdots, 10 $.  SO(10) is defined by the set of real
orthogonal transformations $O_{ij} :  O^T O = 1$ such that $10'_i = O_{ij}
10_j$.  Infinitesimal SO(10) rotations are given by $O = 1 + i \tilde{\omega}$
with  $\tilde{\omega}^T = - \tilde{\omega}$.  We can always express the $10
\times 10$ antisymmetric matrix $\tilde{\omega}$ in the canonical form
$\tilde{\omega}_{ij} \equiv \omega_{ab} \Sigma^{ab}_{ij}$.  $\omega_{ab}$ are
45
real infinitesimal parameters satisfying $\omega_{ab} = - \omega_{ba}$ and
$\Sigma^{ab}_{ij} = i ( \delta^a_i \delta^b_j - \delta^a_j \delta^b_i )$ are
the 45 generators of SO(10) in the 10 dimensional representation.  Note that
the antisymmetric tensor product $(10 \times 10)_A \equiv 45$ is the adjoint
representation.

The SO(10) generators satisfy the Lie algebra
$$ [ \Sigma^{ab}, \; \Sigma^{cd} ]_{ik} \equiv \Sigma^{ab}_{ij}
\Sigma^{cd}_{jk} - \Sigma^{cd}_{ij} \Sigma^{ab}_{jk} = [ \Sigma^{ad}_{ik}
\delta_{bc} - \Sigma^{ac}_{ik} \delta_{bd} + \Sigma^{bc}_{ik} \delta_{ad} -
\Sigma^{bd}_{ik} \delta_{ac} ].$$
The adjoint representation transforms as follows :  $45'_{ij} = O_{ik} O_{jl}
45_{kl}$  or  $ 45'_{ij} = ( O 45 O^T )_{ij}$.

In general the tensor product $(10 \times 10) = (10 \times 10)_A + (10 \times
10)_S = 45 + 54 + 1$.  The 54 dimensional representation is denoted by the
symmetric tensor  $54_{ij} = 54_{ji}, \; Tr(54) = 0$ with transformations $54'
= O 54 O^T$.

The spinor representation of SO(10) can be defined in terms of $2^5 \times 2^5$
dimensional representations of a Clifford algebra $\Gamma_i, \; i = 1, \cdots,
10$ , just as for example the spinor representation of SO(4) is represented in
terms of $4 \times 4$  Dirac gamma matrices (see for example, Georgi, ``Lie
Algebras in Particle Physics" for a more detailed discussion\cite{georgi}).
The
$\Gamma$s satisfy $\Gamma_i^{\dagger} = \Gamma_i, \;\; \{ \Gamma_i, \; \Gamma_j
\} = 2 \delta_{ij}$.  They can explicitly be expressed in terms of tensor
products of 5 Pauli matrices, although we will not do this here.  We can also
define  $\Gamma_{11} \equiv \prod^{10}_{i = 1} \Gamma_i$ satisfying $\{
\Gamma_{11},  \; \Gamma_i \} = 0$ for all i.   The generators of SO(10) in the
spinor representation are now given by $$ \Sigma_{ij} = {i \over 4} [ \Gamma_i,
\; \Gamma_j ] .$$  Note $ [ \Gamma_{11}, \; \Sigma_{ij} ] = 0  $ and
$\Gamma_{11}^2 = 1$.  Hence $ \Gamma_{11}$ has eigenvalues  $\pm 1$ which
divides the 32 dimensional spinor into two irreducible representations of
SO(10)
which are the $16$ and $\overline{16}$ spinor representations.

In order to generate some intuition on
how SO(10) acts on the spinor representations, we use the gamma matrices to
define operators satisfying a Heisenberg algebra of creation and annihilation
operators. Let $$A_{\alpha} = {\Gamma_{2\alpha-1} + i \Gamma_{2\alpha} \over
2}, \;\; \alpha = 1, \cdots, 5$$ and $$ A^{\dagger}_{\alpha} =
{\Gamma_{2\alpha-1} - i \Gamma_{2\alpha} \over 2}. $$  The As satify
$ \{ A_{\alpha} , \; A_{\beta} \} = 0, \;\; \{ A_{\alpha} , \;
A^{\dagger}_{\beta} \} = \delta_{\alpha \beta}.$  We could now rewrite the
generators of SO(10) explicitly in terms of products of $As$ and
$A^{\dagger}s$.  Instead of doing this let me  directly identify an
SU(5) subgroup of SO(10).  In fact the set of generators
$\{ \Sigma_{ij} \}$ are equivalent to the set of generators $\{ Q_a,
\Delta_{\alpha \beta}, \Delta^{\dagger}_{\alpha \beta}, X \}$ defined by
$$Q_a = A^{\dagger}_{\alpha} {\lambda^a_{\alpha \beta} \over 2} A_{\beta}, \;\;
a = 1, \cdots, 24$$ where $\lambda^a_{\alpha \beta}$ are the $5 \times 5$
traceless hermitian  generators of SU(5) in the 5 dimensional representation.
It is then easy to see that the $Qs$ satisfy the Lie algebra of SU(5), $[ Q_a,
\; Q_b ] = i f_{abc} Q_c $. Define $$\Delta_{\alpha \beta} = A_{\alpha}
A_{\beta}
= - \Delta_{\beta \alpha},\;\; \Delta^{\dagger}_{\alpha \beta} =
A^{\dagger}_{\alpha} A^{\dagger}_{\beta} = - \Delta^{\dagger}_{\beta \alpha}.$$
Finally, we define $$ X = -2 \sum^5_{\alpha = 1} (A^{\dagger}_{\alpha}
A_{\alpha}
- {1 \over 2} ) ,$$ the U(1) generator which commutes with the generators of
SU(5).

Let us now define the $16, \; \overline{16}$ representations explicitly.
Consider first the $16$ which contains a $10 + \overline{5} + 1$ under SU(5).
Let $|0\rangle \equiv |{\bf 1}\rangle \equiv [0] $ be the SU(5) invariant state
contained in the $16$, such that  $Q_a |0\rangle \equiv 0$.  It is thus the
vacuum state for the annihilation operators A (i.e. $A_{\alpha} |0\rangle
\equiv
0$), an SU(5) singlet and a zero index tensor under SU(5) transformations
respectively.  We now have $ \Delta^{\dagger}_{\alpha \beta}|0 \rangle = | {\bf
10} \rangle_{\alpha \beta} = [2]$ a 2 index antisymmetric tensor or $10$ under
SU(5). Finally,  $\epsilon^{\alpha \beta \gamma \delta \lambda}
\Delta^{\dagger}_{\alpha \beta} \Delta^{\dagger}_{\gamma \delta} |0\rangle = |
{\bf \overline{5}} \rangle^{\lambda} = [4] $.  Thus, in summary, we have
defined
the $ 16 = 10 +  \overline{5} + 1$ by
$$ |{\bf 1}\rangle = |0\rangle, \;\;  | {\bf 10} \rangle_{\alpha \beta}
= \Delta^{\dagger}_{\alpha \beta}|0 \rangle, \;\; |{\bf \overline{5}}
\rangle^{\lambda} = \epsilon^{\alpha \beta \gamma \delta \lambda}
\Delta^{\dagger}_{\alpha \beta} \Delta^{\dagger}_{\gamma \delta} |0\rangle. $$
Similarly the $\overline{16} =  \overline{10} + 5 + 1$ is defined by
$$ |{\bf 5} \rangle_{\alpha} = A^{\dagger}_{\alpha} | 0 \rangle, \;\; | {\bf
\overline{10}} \rangle^{\delta \rho} = \epsilon^{\alpha \beta \gamma \delta
\rho}
\Delta^{\dagger}_{\alpha \beta} A^{\dagger}_{\gamma} |0\rangle , \;\;
| {\bf 1}\rangle = \epsilon^{\alpha \beta \gamma \delta \rho}
\Delta^{\dagger}_{\alpha \beta} \Delta^{\dagger}_{\gamma \delta}
A^{\dagger}_{\rho} |0\rangle .$$

SO(10) is a rank 5 group, meaning there are 5 U(1) generators in the Cartan
subalgebra.  The 5 generators can be defined as:
$$ \Sigma_{12} = {i \over 4} [ \Gamma_1, \; \Gamma_2] \equiv (A^{\dagger}_1 A_1
-1/2) ,$$
$$ \Sigma_{34} = {i \over 4} [ \Gamma_3, \; \Gamma_4] \equiv (A^{\dagger}_2 A_2
-1/2) ,$$
$$ \Sigma_{56} = {i \over 4} [ \Gamma_5, \; \Gamma_6] \equiv (A^{\dagger}_3 A_3
-1/2) ,$$
$$ \Sigma_{78} = {i \over 4} [ \Gamma_7, \; \Gamma_8] \equiv (A^{\dagger}_4 A_4
-1/2) ,$$
$$ \Sigma_{9 \;10} = {i \over 4} [ \Gamma_9, \; \Gamma_{10}] \equiv
(A^{\dagger}_5 A_5 -1/2) .$$
The first 3 act on color indices and the last two act on weak indices.  Thus
the SU(5) invariant U(1) generator in the 16 dimensional representation is
given by $$ X = - 2 \sum^5_{\alpha = 1} (A^{\dagger}_{\alpha} A_{\alpha} - 1/2)
= - 2 (\Sigma_{12} + \Sigma_{34} + \Sigma_{56} + \Sigma_{78} + \Sigma_{9 10})
.$$

The 10 dimensional representation can be expressed in terms of a $(5 \times 5)
\otimes (2 \times 2)$ tensor product notation.  We can use the above formula to
write an expression for $X$ in this basis.  We find $$  X = 2 x
\otimes \eta $$ where  $$x = \left( \bar{ccccc} 1 & 0 & 0 & 0 & 0 \\ 0 & 1 & 0
&
0 & 0 \\0 & 0 & 1 & 0 & 0 \\0 & 0 & 0 & 1 & 0 \\0 & 0 & 0 & 0 & 1 \ear \right)
$$ and $$ \eta = \left( \bar{cc} 0 & -i \\ i & 0 \ear \right) . $$  Similarly
we
can identify the other U(1)s which commute with $SU(3) \times SU(2)  \times
U(1)_Y$: $$ Y = - {2 \over 3} \sum^3_{\alpha = 1} (A^{\dagger}_{\alpha}
A_{\alpha} - 1/2) + \sum^5_{\alpha = 4} (A^{\dagger}_{\alpha} A_{\alpha} -
1/2)|_{on 16} = y \otimes \eta|_{on 10}$$ where $ y = \left( \bar{ccccc} 2/3 &
0
& 0 & 0 & 0 \\ 0 & 2/3 & 0 & 0 & 0 \\0 & 0 & 2/3 & 0 & 0 \\0 & 0 & 0 & -1 & 0
\\0 & 0 & 0 & 0 & -1 \ear \right) $; $$B - L = - {2 \over 3}  \sum^3_{\alpha =
1}
(A^{\dagger}_{\alpha} A_{\alpha} - 1/2)|_{on 16} = {2 \over 3} (b-l) \otimes
\eta
$$ where $ ( b-l) = \left(  \bar{ccccc} 1 & 0 & 0 & 0 & 0 \\ 0 & 1 & 0 & 0 & 0
\\0 & 0 & 1 & 0 & 0 \\0 & 0 & 0 & 0 & 0 \\0 & 0 & 0 & 0 & 0 \ear \right) $; and
$$ T_{3R} = - {1 \over 2} \sum^5_{\alpha = 4} (A^{\dagger}_{\alpha} A_{\alpha}
-
1/2)|_{on 16} = {1 \over 2} t_{3R} \otimes \eta $$ where $ t_{3R} = \left(
\bar{ccccc} 0 & 0 & 0 & 0 & 0 \\ 0 & 0 & 0 & 0 & 0 \\0 & 0 & 0 & 0 & 0 \\0 & 0
&
0 & 1 & 0 \\0 & 0 & 0 & 0 & 1 \ear \right) $.

It is a useful exercise to use
the definition of the $16$ defined above and the definition of $Y$ in terms of
number operators to identify the hypercharge assignments of the states in the
$16$.

Note that we will use fields in the adjoint (45) representation to break SO(10)
to the SM.  A 45 vev in the $X$ direction will break SO(10) to $ SU(5) \times
U(1)_X$. The vev of a $16 + \overline{16}$ in the $\overline{\nu}$ directions
can then break $X$ leaving SU(5) invariant. We could then use a 45 with vev in
either the $Y, B-L$ or $T_{3R}$ directions to break SU(5) to $SU(3) \times
SU(2)
\times U(1)_Y$.  Note also that either $(X, Y) $ or $(B-L, T_{3R})$ span the 2
dimensional space of U(1)s which commute with $SU(3) \times SU(2) \times
U(1)_Y$.

Finally, the 16 of SO(10) contains one family of fermions and their
supersymmetric partners.  The 10 of SO(10) contains a pair of Higgs doublets
necessary to do the electroweak breaking.  Under SU(5) we have $ 10 = 5 +
\overline{5}$.  The simplest dimension 4 Yukawa coupling of the electroweak
Higgs to a single family (consider the third generation) is given by
$$ A 16_3 \; 10 \; 16_3.$$  The SO(10) symmetry relation which follows is
$$\lambda_t = \lambda_b = \lambda_{\tau} = \lambda_{\nu_{\tau}} = A .$$

\section{The Effective Operator Approach}

We have studied the supersymmetrized Georgi-Jarlskog texture for fermion
masses.  It is remarkably successful in describing the low energy data.
Nevertheless there are some inherent shortcomings with the texture approach.
\begin{enumerate}
\item The texture of zeros is adhoc - perhaps there are others which work
better or have fewer parameters.
\item Down and charged lepton Yukawa matrices are related, but {\em not} up
quarks -- this is so even for the SO(10) version of the theory.
\item There is no explanation of the family hierarchy - the arbitrary
parameters simply satisfy $A >> B >> C$ and $D >> E >> F$.
\item The important Clebsch factor of 3 requires a Higgs 45 dimensional
representation in SU(5) or a $\overline{126}$ in SO(10) -  these are large
representations which make it difficult to construct complete GUT theories.
\end{enumerate}

The third problem above suggests we consider the higher dimension effective
operators of Froggatt-Nielsen\cite{froggatt}.  Combine this with a desire for
maximal predictability  and we are led to consider GUTs with additional family
symmetries.  In the remainder of these lectures we will describe an
effective supersymmetric SO(10) operator analysis of fermion masses. We define
a
procedure for finding the dominant operator set reproducing the low energy
data.
In the minimal operator sets we have just six parameters in the fermion
mass matrices.  We use the six best known low energy parameters as
input to fix these six unknowns and then predict the rest. These
theories are supersymmetric[SUSY] SO(10) grand unified theories[GUTs]\cite{gg}.
In the next two sections I want to briefly motivate these choices.

\newpage

\section{Why SUSY GUTs?}

Looking back at the history of particle physics, it is clear that much of our
understanding comes from using symmetries.  This is because, even without a
complete understanding of the dynamics,  symmetries can be used to relate
different observables.  Here too we want to correlate the known low energy
data,
the three gauge couplings and the fermion masses and mixing angles.  We want to
describe these 16 parameters in terms of fewer fundamental numbers.  GUTs allow
us to do just that. In fact using this symmetry we can express the low energy
data as follows -- $$Observable =  Input\; parameters \times Boundary
\;condition \; at \; M_G \times RG \;factor $$ where the {\it observable} is
the
particular low energy data we want to calculate,  the {\it input parameters} is
the set of fundamental parameters defined at the GUT scale and the last factor
takes into account the renormalization group running of the experimental
observable from $M_G$ to the low energy scale. The grand unified symmetry SU(5)
(or SO(10), E(6) etc.) determines the {\it boundary conditions at
$M_G$}\cite{gqw}. These are given in terms of Clebsch-Gordan coefficients
relating different observables.  Of course,  these relations are only valid at
the GUT scale and the RG equations are necessary to relate them to experiment.
It is through the RG equations that supersymmetry enters.  We will assume that
only the states present in the minimal supersymmetric standard model[MSSM] are
in the theory below $M_G$.  We assume this because it works. Consider the GUT
expression for the gauge  couplings -- $$ \alpha_i(M_Z) = \alpha_G \;
R_i(\alpha_G, {M_G \over M_Z})$$ where the boundary condition is $R_i(\alpha_G,
1) \equiv 1 + \cdots$.  The input parameters are $\alpha_G$ and $M_G$ and the
Clebschs in this case are all one.  Thus we obtain the well known result that
given $\alpha$ and $\sin^2\theta_W$ measured at $M_Z$ we predict the value for
$\alpha_s(M_Z)$\cite{drw1}(For recent analysis of the data, see \cite{recent}).
Note that SUSY without GUTs makes no prediction, since there is no symmetry to
specify the boundary conditions and GUTs without SUSY makes the wrong
prediction.

I should also point out that the SO(10) operator analysis for fermion masses
that I am about to describe is not new. This analysis was carried out 10 years
ago with the result that the favored value of the top quark mass was about 35
GeV\cite{pre-adhrs}.

\section{Why SO(10)?}

There are two reasons for using SO(10).

\begin{enumerate}
\item  It is the smallest group in which all the fermions in one family
fit into one irreducible representation, i.e. the ${\bf 16}$.  Only one
additional state needs to be added to complete the multiplet and that
is a right-handed neutrino.  In larger gauge groups,  more as yet
unobserved states must be introduced to obtain complete multiplets.
Thus we take ${\bf 16}_i \supset \{ U_i, D_i, E_i, \nu_i \}, i = 1,2,3$
for the 3 families with the third family taken to be the heaviest.
{\em Since SO(10) Clebschs can now relate  $U, D, E$ and $\nu$ mass
matrices, we can in principle reduce the number of fundamental
parameters in the fermion sector of the theory.}  We return to this
point below.
\item  In any SUSY theory there are necessarily two higgs doublets --
$H_u$ and $H_d$.  Both these states fit into the ${\bf 10}$ of SO(10)
and thus their couplings to up and down type fermions are also given by
a Clebsch.  There are however six additional states in the ${\bf 10}$
which transform as a ${\bf 3}$ + ${\bf \overline{3}}$ under color.
These states contribute to proton decay and must thus be heavy.  The
problem of giving these color triplet states large mass of order $M_G$
while keeping the doublets light is sometimes called the second gauge
hierarchy problem.  This problem has a natural solution in SO(10) which
we discuss later\cite{dw}.
\end{enumerate}

Note that the gauge group SO(10) has to be spontaneously broken to the
gauge group of the SM -- $SU(3)\times SU(2)\times U(1)$.  This GUT
scale breaking can be accomplished by a set of states including \{
${\bf 45, 16, \overline{16}, \cdots }$\}.   The ${\bf 45}$(the adjoint
representation) enters into our construction of effective fermion mass
operators, thus I will discuss it in more detail in the next section.

I promised to return to the possibility of reducing the number of
fundamental parameters in the fermion sector of the theory.  Recall
that there are 13 such parameters.  Using symmetry arguments we can now
express the matrices ${\bf D, E},$ and ${\bf \nu}$ in terms of one
complex 3x3 matrix, $U$.  Unfortunately, this is not sufficient to
solve our problem.  There are 18 arbitrary parameters in this one
matrix.  In order to reduce the number of fundamental parameters we
must have zeros in this matrix.  We thus need new {\it family
symmetries} to enforce these zeros.

\section{The Big Picture}

Let us consider the big picture(see Fig. 10).  Our low energy observer
measures the physics at the electroweak scale and perhaps an order of
magnitude above.  Once the SUSY threshold is crossed we have direct
access to the effective theory at $M_G$, the scale where the 3 gauge
couplings meet.  Of course the GUT scale $M_G \sim 10^{16}$ GeV is
still one or two orders of magnitude below some more fundamental scale
such as the Planck or string scales (which we shall refer to as M).
Between M and $M_G$ there may be some substructure.  In fact, we may be
able to infer this substructure by studying fermion masses.

In our analysis we assume that the theory below the scale M is
described by a SUSY SO(10) GUT.  Between $M_G$ and $M$, at a scale
$v_{10}$, we assume that the gauge group SO(10) is broken spontaneously
to SU(5).  This can occur due to the vacuum expectation value of an
adjoint scalar in the X direction  and  the expectation values of a 16 and a
$\overline{16}$(denoted by $\Psi$ and $\overline{\Psi}$ respectively).
Then SU(5) is broken at the scale $v_5 = M_G$ to the SM gauge group.
This latter breaking can be done by different adjoints (45) in the Y,
B-L or T$_{3R}$ directions.

{\em Why consider 4 particular breaking directions for the 45 and no
others?}
The X and Y directions are orthogonal and span the two dimensional
space of U(1) subgroups of SO(10) which commute with the SM.  B-L and
T$_{3R}$ are also orthogonal and they span the same subspace.
Nevertheless we consider these four possible breaking directions and
these are the {\em only directions} which will enter the effective
operators for fermion masses.  Why not allow the X and Y directions or
any continous rotation of them in this 2d subspace of U(1) directions .
The answer is that there are good dynamical arguments for assuming that
these and only these directions are important. The X direction breaks
SO(10) to an intermediate SU(5) subgroup and it is reasonable to assume
that this occurs at a scale $v_{10} \ge v_5$.  Whether $v_{10}$ is
greater than $v_5$ or equal will be determined by the LED.  The B-L
direction is required for other reasons.  Recall the color triplet
higgs in the 10 which must necessarily receive large mass.  As shown by
Dimopoulos and Wilczek\cite{dw},  this doublet-triplet splitting can
naturally occur by introducing a 10 45 10 type coupling in the
superspace potential.  Note that the higgs triplets carry non-vanishing
B-L charge while the doublets carry zero charge.  Thus when the 45 gets
a vacuum expectation value[vev] in the B-L direction it will give mass
to the color triplet higgs at $v_5$ and leave the doublets massless.
Thus in any SO(10) model which solves this second hierarchy problem,
there must be a 45 pointing in the B-L direction.  We thus allow for
all 4 possible breaking vevs  --- X, Y, B-L and T$_{3R}$.  Furthermore
we believe this choice is ``natural" since we know how to construct
theories which have these directions as vacua without having to tune
any parameters.

Our fermion mass operators have dimension $\ge 4$.  {\em From where
would these higher dimension operators come?} Note that by measuring
the LED we directly probe the physics in some effective theory at
$M_G$.  This effective theory can, and likely will, include operators
with dimension greater than 4.     Consider, for example, our big
picture looking down from above.  String theories are very fundamental.
They can in principle describe physics at all scales.  Given a
particular string vacuum,  one can obtain an effective field theory
valid below the string scale M.  The massless sector can include the
gauge bosons of SO(10) with scalars in the 10, 45 or even 54
dimensional representations.  In addition, we require 3 families of
fermions in the 16.  Of course, in a string context when one says that
there are 3 families of fermions what is typically meant is that there
are 3 more 16s than $\overline{16}$s.  The extra 16 + $\overline{16}$
pairs are assumed to get mass at a scale $\ge M_G$,  since there is no
symmetry which prevents this.  When these states are integrated out in
order to define the effective field theory valid below $M_G$ they will
typically generate higher dimension operators.

Consider a simple two family model.  Let $16_2,\; 16_3$ represent the  2
heaviest families of quarks and leptons,  $\Psi_i,\; \overline{\Psi}_i,\;\; i =
1,2$ are heavy $16, \overline{16}$ states with mass of order $M_G$, $A_2,
\tilde{A}$ are in the 45 dimensional representation and $10$ contains the
electroweak Higgs doublets.  In this example we have 4 $16s$ and 2
$\overline{16}s$. At the scale M we assume the superspace potential has the
form
$$16_3 10 16_3 + g_3 16_3 A_2 \overline{\Psi}_1 + \tilde{g}_1 \overline{\Psi}_1
\tilde{A} \Psi_1 + g_2 16_2 A_2 \overline{\Psi}_2 + \tilde{g}_2
\overline{\Psi}_2 \tilde{A} \Psi_2 + \Psi_1 10 \Psi_2.$$  We now assume that
$\langle \tilde{A} \rangle \sim X$ and $ \langle A_2 \rangle \sim Y$ with
$\langle \tilde{A} \rangle  >> \langle A_2 \rangle$.  Thus the dominant
contribution to the mass of the states  $\Psi_i,\; \overline{\Psi}_i,\;\; i =
1,2$ is given by $ \tilde{g}_i \langle \tilde{A} \rangle$.  In order to define
the effective theory at $M_G$, we  must integrate these states out of the
theory.  As a result we obtain the effective mass operators -
$$O_{33} = 16_3^p 10^{-2p} 16_3^p, \;\;\; O_{32} = 16_3^p \left({A_2 \over
\tilde{A}} \right)^{q_2 - t} 10^{-2p} \left({A_2 \over
\tilde{A}} \right)^{q_2 - t} 16_2^{p-2q_2+2t}$$ which can be read off the tree
diagrams in fig. 11.   The superscripts in this formula denote independent U(1)
charges which may be assigned to the fields.  The sum of the charges at any
vertex must vanish for $U(1)_p, U(1)_{q_2}, U(1)_t$ to be symmetries of the
theory.  Note, at the level of the effective operators,  the operator
$$ 16_3^p \left({A_2 \over \tilde{A}} \right)^{2q_2 - 2t} 10^{-2p} 16_2^{p -
2q_2+2t} $$ also preserves all 3 U(1) symmetries.  This operator is not
equivalent to $O_{32}$ above. It cannot be obtained however by integrating out
the heavy fields.  Thus the symmetries of the full theory restrict the order of
operators appearing in the effective theory.

The operators $O_{33}$ and $O_{32}$ represent only the first term in a power
series in the ratio $ \left|{\langle A_2 \rangle \over \langle \tilde{A}
\rangle}\right|^2$.  We can obtain the complete effective theory by
diagonalizing
the $4 \times 2$ mass matrix
$$ \bar{cc}  & \bar{cllcr}  & \Psi_1 & \Psi_2 & 16_3 & 16_2 \ear \\
 \bar{c} \overline{\Psi}_1 \\ \overline{\Psi}_2 \ear & \left( \bar{cccc}
\tilde{g}_1 \langle \tilde{A}\rangle & 0 & g_3 \langle A_2 \rangle & 0 \\ 0 &
\tilde{g}_2 \langle \tilde{A} \rangle & 0 & g_2 \langle A_2 \rangle \ear
\right) \ear. $$   The mass eigenstates are given by
$$\Psi'_1 = (\tilde{g}_1 \langle \tilde{A} \rangle \Psi_1 + g_3 \langle A_2
\rangle 16_3 )/m_1,$$
$$16'_3 = ( -g_3 \langle A_2 \rangle \Psi_1 + \tilde{g}_1 \langle \tilde{A}
\rangle 16_3 )/m_1$$ where $m_1 = \sqrt{\tilde{g}_1^2 |\langle \tilde{A}
\rangle|^2 + g_3^2 |\langle A_2 \rangle|^2}$.  Similarly,
$$\Psi'_2 = (\tilde{g}_2 \langle \tilde{A} \rangle \Psi_2 + g_2 \langle A_2
\rangle 16_2 )/m_2,$$
$$16'_2 = ( -g_2 \langle A_2 \rangle \Psi_2 + \tilde{g}_2 \langle \tilde{A}
\rangle 16_2 )/m_2$$ where $m_2 = \sqrt{\tilde{g}_2^2 |\langle \tilde{A}
\rangle|^2 + g_2^2 |\langle A_2 \rangle|^2}$.  The states $16'_3, 16'_2$ are
massless,  while  the other states have mass terms $ \sum_{i = 1}^2 (m_i
\overline{\Psi}_i \Psi'_i)$.  We can now invert the relations to get
$$16_3 = (\tilde{g}_1 \langle \tilde{A} \rangle 16'_3 + g_3 \langle A_2 \rangle
\Psi'_1)/m_1,$$
$$\Psi_1 = (\tilde{g}_1 \langle \tilde{A} \rangle \Psi'_1 - g_3 \langle A_2
\rangle 16'_3)/m_1,$$
$$\Psi_2 = (\tilde{g}_2 \langle \tilde{A} \rangle \Psi'_2 - g_2 \langle A_2
\rangle 16'_2)/m_2.$$  The effective field theory is now obtained by taking
the terms in the superspace potential $16_3 10 16_3 + \Psi_1 10 \Psi_2$ and
replacing $16_3, \Psi_1, \Psi_2$ by their massless components.  We find
$$16_3 \left({1 \over \sqrt{1 + \left|{ g_3 \langle A_2 \rangle \over
\tilde{g}_1 \langle \tilde{A} \rangle}\right|^2}}\right) 10 \left({1 \over
\sqrt{1 + \left|{ g_3 \langle A_2 \rangle \over \tilde{g}_1 \langle \tilde{A}
\rangle}\right|^2}}\right) 16_3  $$
$$+ 16_3 \left({ g_3 \langle A_2 \rangle \over
\tilde{g}_1 \langle \tilde{A} \rangle} \right)\left({1 \over \sqrt{1 + \left|{
g_3 \langle A_2 \rangle \over \tilde{g}_1 \langle \tilde{A}
\rangle}\right|^2}}\right) 10 \left({ g_2 \langle A_2 \rangle \over \tilde{g}_2
\langle \tilde{A} \rangle} \right)\left({1 \over \sqrt{1 + \left|{ g_2 \langle
A_2 \rangle \over \tilde{g}_2 \langle \tilde{A} \rangle}\right|^2}}\right)
16_2.
$$

\section{Operator Basis for Fermion Masses at $M_G$}

Let us now consider the general {\bf operator basis for fermion
masses}.  We include operators of the form
$$
{\bf O_{ij}} = {\bf 16_i} ~( \cdots )_n ~{\bf 10} ~( \cdots )_m ~{\bf
16_j}  $$
where
$$( \cdots )_n =  {M_G^k ~45_{k+1} \cdots 45_n \over M_P^l
{}~45_X^{n-l}}
$$
and the $45$ vevs in the numerator can be in any of the 4 directions,
${\bf X, Y, B-L, T_{3R}}$ discussed earlier.

It is trivial to evaluate the
Clebsch-Gordon coefficients associated with any particular operator
since the matrices $X,Y,B-L,T_{3R}$ are diagonal.  Their eigenvalues on
the fermion states are given in Table 2.

\begin{table}[t]
\begin{center}
\begin{tabular}{|c|c|c|c|c|}
\multicolumn{5}{l}{Table~2. Quantum numbers of the}\\
 \multicolumn{5}{l}{four 45 vevs on fermion states.}  \\
\multicolumn{5}{l}{Note, if $u$ denotes a left-handed}\\
\multicolumn{5}{l}{up quark, then ${\ou}$ denotes } \\
\multicolumn{5}{l}{a left-handed charge conjugate }\\
\multicolumn{5}{l}{up quark. }\\
\multicolumn{1}{c}{}&\multicolumn{4}{c}{}\\ \hline\hline
& ${\bf X}$ & ${\bf Y}$ & ${\bf B-L}$ & ${\bf T_{3R}}$
\\ \hline $u$ &  1 &  1/3 & 1 & 0  \\
${\ou}$ &  1 & -4/3 & -1 & -1/2 \\
$d$ & 1 & 1/3 & 1 & 0 \\
${\od}$ & -3 & 2/3 & -1 & 1/2 \\
$e$ & -3 & -1 & -3 & 0 \\
${\oe}$ & 1 & 2 & 3 & 1/2 \\
$\nu$ & -3 & -1 & -3 & 0 \\
${\onu}$ & 5 & 0 & 3 & -1/2\\ \hline
\end{tabular}
\end{center}
\end{table}

It is probably useful to consider a couple of examples of effective operators
and work out their contributions to fermion mass matrices before we continue
with our discussion of the systematic search over all operator sets which are
consistent with the low energy data.  For our first example consider the 2
family effective theory discussed earlier.   The superspace potential is given
by $$O_{33}+ O_{32} = 16_3 10 16_3 + 16_3 \left({ \langle A_2 \rangle
\over \langle \tilde{A} \rangle} \right) 10 \left({\langle A_2\rangle \over
\langle \tilde{A} \rangle} \right) 16_2 .$$ We now assume  $\langle A_2 \rangle
= a_2 e^{i\alpha_2} Y, \;\; \langle \tilde{A} \rangle = \tilde{a} e^{i
\tilde{\alpha}} X$ with $a_2 \sim M_G$ and $\tilde{a} = v_{10} > M_G$.

We can now evaluate the Yukawa matrices.  We find
$${\cal U} = \left(\bar{cc} 0 & x'_u B \\ x_u B & A \ear \right),$$
$${\cal D} = \left(\bar{cc} 0 & x'_d B \\ x_d B & A \ear \right),$$
$${\cal E} = \left(\bar{cc} 0 & x'_e B \\ x_e B & A \ear \right).$$
The constant $B$ is given in terms of ratio of scales $$B = ({\rm ratio
\;\;of\;\; coupling\;\; constants}) \left({a_2 \over \tilde{a}}\right)^2  $$
where we have explicitly redefined the phases of fermions to remove the
arbitrary
phase.  Finally we evaluate the Clebschs $$ x_u = x'_u = -4/9,  x_d = x'_d =
-2/27,  x_e = x'_e = 2/3 .$$  A particular ratio of Clebschs  $$\chi \equiv
{|x_u - x_d| \over \sqrt{|x_u x'_u|}} = 5/6 .$$ In this case the Yukawa
matrices
are symmetric.

In the next example, replace the operator $O_{32}$ above by
$$O_{32} = 16_3 {A_1 \over \tilde{A}} 10 {A_2 \over \tilde{A}} 16_2 $$
where $ \langle A_1 \rangle = a_1 e^{i \alpha_1} (B-L)$.  In this case $B
\approx \left({a_1 a_2 \over \tilde{a}^2}\right)$.   We find the Clebschs
$$x_u = -1/3, x'_u = -4/3,  x_d = 1/9,  x'_d = -2/9,  x_e = 1, x'_e =
2 $$  and in this case $$\chi \equiv {|x_u - x_d| \over \sqrt{|x_u x'_u|}} =
2/3
.$$  You see that the Yukawa matrices are no longer symmetric.

\section{Dynamic Principles}

Now consider the dynamical principles which guide us towards a
theory of fermion masses.
\begin{description}
\item[0.] At zeroth order, we work in the context of a SUSY GUT with
the MSSM  below  $M_G$.
\item[1.] We use SO(10) as the GUT symmetry with three families of
fermions $\{ 16_i  ~~i = 1,2,3 \}$ and the minimal electroweak Higgs
content in one $10$.  SO(10) symmetry relations allow us to
reduce the number of fundamental parameters.
\item[2.] We assume that there are also family symmetries which enforce
zeros of the mass matrix,  although we will not specify these
symmetries at this time.  As we will make clear in section 12, these
symmetries will be realized at the level of the fundamental theory
defined below $M$.
\item[3.] Only the third generation obtains mass via
a dimension 4 operator.  The fermionic sector of the Lagrangian thus
contains the term $ A ~O_{33} \equiv A ~~16_3 ~10 {}~16_3$.  This term
gives mass to  t, b and $\tau$.  It results in the
symmetry relation --- $\lambda_t = \lambda_b = \lambda_{\tau} \equiv A$
at $M_G$.   This relation has been studied before by Ananthanarayan,
Lazarides and Shafi\cite{als} and using $m_b$ and $m_{\tau}$ as
input it leads to reasonable results for $m_t$ and $\tan \beta$.
\item[4.] All other masses come from operators with dimension $> 4$.
As a consequence,  the family hierarchy is related to the ratio
of scales above $M_G$.

\item[5.]  [{\bf Predictivity requirement}] ~We demand the
\underline{minimal set} of effective fermion mass operators at $M_G$
\underline{consistent with the {\bf LED}}.
\end{description}

\section{Systematic Search}

Our goal is to find the {\em minimal} set of fermion mass
operators consistent with the LED.  With any given operator set
one can evaluate the fermion mass matrices for up and down quarks and
charged leptons.  One obtains relations between mixing angles and
ratios of fermion masses which can be compared with the data.  It is
easy to show, however, without any detailed calculations that the
minimal operator set consistent with the LED is given by
\begin{eqnarray}
 & O_{33} + O_{23} + O_{22} + O_{12}&  ---
``22" ~{\rm texture} \nonumber\\
 {\rm or} & &  \nonumber\\
&  O_{33} + O_{23} + O'_{23} + O_{12}& --- ``23'" ~{\rm texture}
\nonumber
\end{eqnarray}

It is clear that at least 3 operators are needed to give
non-vanishing and unequal masses to all charged fermions, i.e. $ ~det
(m_a)  \neq 0$ for $a = u,d,e$.  That the operators must be in the [33,
23 and 12]
slots is not as obvious but is not difficult to show.  It is then
easy to show that 4 operators are required in order to have  CP
violation.  This is because, with only 3 SO(10) invariant operators,
we can redefine the phases of the three 16s of fermions to remove the
three arbitrary phases.  With one more operator, there is one
additional phase which cannot be removed.  A corollary of this
observation is that this minimal operator set results in just 5
arbitrary parameters in the Yukawa matrices of all fermions,  4
magnitudes and one phase\footnote{This is two fewer parameters than was
necessary in our previous analysis (see \cite{dhr})}.  This is the
minimal parameter set which can be obtained without solving the
remaining problems of the fermion mass hierarchy, one overall real
mixing angle and a CP violating
phase.  We should point out however that the problem of understanding
the fermion mass hierarchy and mixing has been rephrased as the problem
of understanding the hierarchy of scales above $M_G$.

{}From now on I will just consider models with ``22" texture.  This is
because they can reproduce the observed hierarchy of fermion masses
without fine-tuning\footnote{For more details on this point, see
section 14 below or refer to \cite{adhrs}.}.  Models with ``22" texture
give the following Yukawa matrices at $M_G$ (with electroweak doublet
fields on the right) --

$$
{\bf \lambda_a} = \left( \begin{array}{ccc}
0 & z'_a ~C & 0\\
z_a ~C & y_a ~E ~e^{i \phi} & x'_a ~B\\
0 & x_a ~B & A
\end{array} \right) $$
 with the subscript $a = \{ u, d, e\}$.  The constants
$x_a, x'_a, y_a, z_a, z'_a$ are Clebschs which can be determined once
the 3 operators ( $O_{23},O_{22}, O_{12}$) are specified.  Recall, we
have taken $O_{33} = A ~16_3 ~10 ~16_3$, which is why the Clebsch in
the 33 term is independent of $a$. Finally, combining the Yukawa
matrices with the Higgs vevs to find the fermion mass matrices we have
6 arbitrary parameters given by $A, B, C, E, \phi$ and $\tan \beta$
describing 14 observables.  We thus obtain 8 predictions.  We shall
use the best known parameters, $m_e, m_{\mu}, m_{\tau}, m_c, m_b,
|V_{cd}|$, as
input to fix the 6 unknowns.  We then predict the values of $m_u, m_d,
m_s,
m_t, \tan \beta, |V_{cb}|, |V_{ub}|$ and $J$.

Note: since the predictions are correlated, our analysis would be much
improved if we minimized some $\chi^2$ distribution and obtained a best
fit to the data.  Unfortunately this has not yet been done.  In the
paper however we do include some tables (see for example Table 5) which
give all the predictions for a particular set of input parameters.

\section{Results}

The results for the 3rd generation are given in fig. 12.  Note that
since the parameter A is much bigger than the others we can essentially
treat the 3rd generation independently.  The small corrections, of
order $(B/A)^2$, are however included in the complete analysis. We find
the pole mass for the top quark $M_t = 180 \pm 15$ GeV and $\tan \beta
= 56\pm 6$ where the uncertainties result from variations of our input
values of the $\overline{MS}$ running mass $m_b(m_b) = 4.25 \pm 0.15$
and $\alpha_s(M_Z)$ taking values $.110 - .126$. We used two loop RG
equations for the MSSM from $M_G$ to $M_{SUSY}$;  introduced a
universal SUSY threshold at $M_{SUSY} = 180$ GeV with 3 loop QCD and 2
loop QED RG equations below $M_{SUSY}$.  The variation in the value of
$\alpha_s$ was included to indicate the sensitivity of our results to
threshold corrections which are necessarily present at the weak and GUT
scales.  In particular, we chose to vary $\alpha_s(M_Z)$ by letting
$\alpha_3(M_G)$ take on slightly different values than $\alpha_1(M_G) =
\alpha_2(M_G) = \alpha_G$.

The following set of operators passed a straightforward but coarse
grained search discussed in detail in the paper\cite{adhrs}.  They
include
the diagonal dimension four coupling of the third generation --

\begin{eqnarray}
    O_{33} = &  16_3\ 10\ 16_3 &  \nonumber
\end{eqnarray}
{}.

The six possible $O_{22}$ operators --

\begin{eqnarray}
O_{22}  =  &  &  \nonumber \\
& 16_2 \ {45_X \over M} \ 10 \ {45_{B-L} \over 45_X} \ 16_2 & (a) \nonumber\\
& 16_2 \ {M_G \over \ 45_X} \ 10 \ {45_{B-L}\over M} \ 16_2 &(b) \nonumber\\
& 16_2 \ {45_X \over M} \ 10 \ {45_{B-L} \over M} \ 16_2 & (c) \nonumber\\
& 16_2 \ 10 \ {45_{B-L}\over 45_X} \  16_2 &(d) \nonumber \\
& 16_2 \ 10 \ {45_X \ 45_{B-L} \over M^2} \ 16_2  & (e) \nonumber\\
& 16_2 \ 10 \ {45_{B-L} \ M_G \over 45_X^2}\ 16_2 & (f) \nonumber
\end{eqnarray}
Note: in all cases the Clebschs $y_i$ ( defined by $O_{22}$ above)
satisfy

$$
y_u : y_d : y_e = 0 : 1 : 3.
$$
This is the form familiar from the Georgi-Jarlskog texture\cite{gj}.
Thus all six of these operators lead to {\it identical} low energy
predictions.

Finally there is a unique operator $O_{12}$ consistent with the LED --
\begin{eqnarray}
O_{12} = & 16_1 \left( {45_X\over M}\right)^3 \ 10 \left( {45_X \over
M}\right)^3
16_2 &  \nonumber
\end{eqnarray}

The operator $O_{23}$  determines the KM element $V_{cb}$ by the
relation --
$$
V_{cb} = \chi \ \sqrt{ {m_c \over m_t}} \times ( RG factors)
$$
where the Clebsch combination $\chi$ is given by
$$
\chi \equiv {|x_u-x_d|\over \sqrt{|x_ux_u'|}}
$$
$m_c$ is input, $m_t$ has already been determined and the RG factors are
calculable.  Demanding the   experimental constraint
$V_{cb} < .054$ we find the constraint $\chi < 1$.
A search of all operators of dimension 5 and 6 results in the 9
operators given below. Note that there only three different values of
$\chi = 2/3, ~5/6,~8/9$ --

\begin{eqnarray}
O_{23} = &  &  \nonumber \\
 & & \chi = 2/3 \nonumber\\
(1)& 16_2 \ {45_{Y} \over M} \ 10 {M_G \over 45_X} \ 16_3 & \nonumber\\
(2)& 16_2 \ {45_{Y} \over M} \ 10\ {45_{B-L}\over 45_X} \ 16_3 & \nonumber\\
(3)& 16_2 \ {45_{Y}\over 45_X} \ 10 \ {M_G \over 45_X} \ 16_3 & \nonumber\\
(4)& 16_2 \ {45_{Y}\over 45_X} \ 10 \ {45_{B-L}\over45_X}   16_3 & \nonumber
\end{eqnarray}
\begin{eqnarray}
& & \chi = 5/6 \nonumber\\
(5)&  16_2 \ {45_{Y} \over M} \ 10 \ {45_{Y}\over 45_X} \ 16_3 & \nonumber\\
(6)&  16_2 \ {45_{Y}\over 45_X} \ 10 \ {45_{Y}\over 45_X} \ 16_3 & \nonumber
\end{eqnarray}
\begin{eqnarray}
& & \chi= 8/9 \nonumber\\
(7)&  16_2 \ 10 \ {M_G^2 \over 45_X^2} \ 16_3 & \nonumber\\
(8)&  16_2 \ 10 \ {45_{B-L} M_G \over 45_X^2} \ 16_3 & \nonumber\\
(9)&  16_2 \ 10 \ {45_{B-L}^2\over 45^2_X} \ 16_3  & \nonumber
\end{eqnarray}
We label the operators (1) - (9), and we use these numbers also to
denote the corresponding models.  {\em Note, all the operators have the
vev $45_X$ in the denominator.  This can only occur if  $v_{10} > M_G$.}

At this point, there are no more simple criteria to reduce the number
of models further.  We have thus performed a numerical RG analysis on
each of the 9 models (represented by the 9 distinct operators $O_{23}$
with their calculable Clebschs $x_a, x'_a, a= u,d,e$  along with the
unique set of Clebschs determined by the operators $O_{33}, O_{22}$ and
$O_{12}$). We then iteratively fit the 6 arbitrary parameters to the
six low energy inputs and evaluate the predictions for each model as a
function of the input parameters.  The results of this analysis are
given in figs. (13 - 19).

Let me make a few comments.  Light quark masses (u,d,s) are
$\overline{MS}$ masses evaluated at 1 GeV while heavy quark masses
(c,b) are evaluated at ($m_c, m_b$) respectively.  Finally, the top
quark mass in fig. 12 is the pole mass.  figs. 13 and 14 are self evident.
In fig. 15, we show the correlations for two of our predictions.  The
ellipse in the $m_s/m_d$ vs. $m_u/m_d$ plane is the allowed region from
chiral Lagrangian analysis\cite{chiral}. One sees that we favor lower
values of $\alpha_s(M_Z)$. For each fixed value of $\alpha_s(M_Z)$,
there are 5 vertical line segments in the $V_{cb}$ vs. $m_u/m_d$ plane.
Each vertical line segment represents a range of values for $m_c$ (with
$m_c$ increasing moving up) and the different line segments represent
different values of $m_b$ (with $m_b$ increasing moving to the left).
In Figure 18 we test our agreement with the observed CP violation in the
K system.  The experimentally determined value of $\epsilon_K = 2.26
\times 10^{-3}$. Theoretically it is given by an expression of the form
$B_K \times \{ m_t, V_{ts}, \cdots \}$.   $B_K$ is the so-called Bag
constant which has been determined by lattice calculations to be in the
range $B_K = .7 \pm .2$\cite{bag}.  In fig. 18 we have used our
predictions for fermion masses and mixing angles as input, along with
the experimental value for $\epsilon_K$, and fixed $B_K$ for the 9
different models.  One sees that model 4 is inconsistent with the
lattice data.  In fig. 19 we present the predictions for each model,
for the CP violating angles which can be measured in B decays.  The
interior of the ``whale" is the range of parameters consistent with the
SM found by Nir and Sarid\cite{ns} and the error bars represent the
accuracy expected from a B factory.

Note that model 4 appears to give too little CP violation and model 9
has uncomfortably large values of $V_{cb}$.  Thus these models are
presently disfavored by the data.  I will thus focus on model 6 in the
rest of these lectures.

\section{Summary}

We have performed a systematic operator analysis of fermion masses in
an effective SUSY SO(10) GUT.  We use the LED to lead us to the theory.
Presently there are 3 models (models 4, 6 \& 9) with ``22" texture
which agree best with the LED, although as mentioned above model 6 is
favored.  In all cases we used the values of $\alpha$ and
$\sin^2\theta_W$ (modulo threshold corrections) to fix
$\alpha_s(M_Z)$.

Table 3 shows the virtue of the ``22" texture.  In the first column are
the four operators.  In the 2nd and 3rd columns are the parameters in
the mass matrix relevant for that particular operator and the input
parameters which are used to fix these parameters.  Finally the 4th
column contains the predictions obtained at each level.  One sees that
each family is most sensitive to a different operator\footnote{This
property is not true of ``23" textures.}.

\begin{table}[t]
\begin{center}
\begin{tabular}{|c|c|c|c|}
\multicolumn{4}{l}{Table~3. Virtue of ``22" texture.}\\
 \multicolumn{4}{c}{}\\ \hline\hline
Operator  & Parameters & Input & Predictions \\ \hline

$O_{33}$ &  $\tan\beta$ \ A &  b \ $\tau$ & t  \ $\tan\beta$  \\
$O_{23}$ &  B & c & $V_{cb}$  \\
$O_{22}$ & E & $\mu$ & s  \\
$O_{12}$ & C \ $\phi$  & e \ $V_{us}$ & u \ d \ ${V_{ub} \over V_{cb}}$
\ J \\ \hline
\end{tabular}
\end{center}
\end{table}

{\em Consider the theoretical uncertainties inherent in our analysis.}
\begin{enumerate}
\item The experimentally determined values of $m_b, m_c$, and
$\alpha_s(M_Z)$ are all subject to strong interaction uncertainties of
QCD.  In addition, the predicted value of $\alpha_s(M_Z)$ from GUTs is
subject to threshold corrections at $M_W$  which can only be calculated
once the SUSY spectrum is known and at $M_G$ which requires knowledge
of the theory above $M_G$.  We have included these uncertainties
(albeit crudely) explicitly in our analysis.

\item In the large $\tan \beta$ regime in which we work there may be
large SUSY loop corrections which will affect our results.  The finite
corrections to the $b$ and $\tau$ Yukawa couplings have been
evaluated\cite{hrs,copw}.  They depend on ratios of soft SUSY breaking
parameters and are significant in certain regions of parameter
space\footnote{There is a small range of parameter space in which our
results are unchanged\cite{hrs}.  This requires threshold corrections
at $M_G$ which distinguish the two Higgs scalars.}.  In particular it
has been shown that the top quark mass can be reduced by as much as
30\%.  Note that although the prediction of fig. 3 may no longer be
valid,  there is still necessarily a prediction for the top quark mass.
It is now however sensitive to the details of the sparticle spectrum
and to the process of radiative electroweak symmetry
breaking\cite{resb}.  This means that the observed top quark mass can
now be used to set limits on the sparticle spectrum.   This analysis
has not been done.  Moreover,  there are also similar corrections to
the Yukawa couplings for the $s$ and  $d$ quarks and for $e$ and $\mu$.
These corrections are expected to affect the predictions for $V_{cb},
m_s, m_u, m_d$.   It will be interesting to see the results of this
analysis.

\item The top, bottom and $\tau$ Yukawa couplings can receive threshold
corrections at $M_G$.  We have not studied the sensitivity to these
corrections.

\item Other operators could in principle be added to our effective
theory at $M_G$.  They might have a dynamical origin. We have assumed
that, if there, they are subdominant.  Two different origins for these
operators can be imagined.  The first is field theoretic. The operators
we use would only be the leading terms in a power series expansion when
defining an effective theory at $M_G$ by integrating out heavier
states.  The corrections to these operators are expected to be about
10\%.   We may also be sensitive to what has commonly been referred to
as Planck slop\cite{pslop}, operators suppressed by some power of the
Planck (or string) scale M.   In fact the operator $O_{12}$ may be
thought of as such.  The question is why aren't our results for the
first and perhaps the second generation,  hopelessly sensitive to this
unknown physics?  This question will be addressed in the next section.
\end{enumerate}

\section{Where are we going?}

In the first half of Table 4 I give a brief summary of the good and bad
features of the effective SUSY GUT  discussed earlier.  Several models
were found with just four operators at $M_G$ which successfully fit the
low energy data. If we add up all the necessary parameters needed in
these models we find just 12.  This should be compared to the SM with
18 or the MSSM with 21.  Thus these theories,  minimal effective SUSY
GUTs[MESG], are doing quite well.  Of course the bad features of the
MESG is that it is not a fundamental theory.  In particular there are
no symmetries which prevent additional higher dimension operators to
spoil our results.  Neither are we able to calculate threshold
corrections, even in principle, at $M_G$.

It is for these reasons that we need to be able to take the MESG which
best describes the LED and use it to define an effective field theory
valid at scales $\le M$.  The good and bad features of the resulting
theory are listed in the second half of Table 4.

\begin{itemize}
\item In the effective field theory below $M$ we must incorporate the
{\it symmetries} which guarantee that we reproduce the MESG with no
additional operators\footnote{This statement excludes the unavoidable
higher order field theoretic corrections to the MESG which are, in
principle, calculable.}

{\em Moreover, the necessary combination of discrete, U(1) or R
symmetries may be powerful enough to restrict the appearance of Planck
slop.}

\item Finally,  the {\it GUT symmetry breaking} sector must resolve the
problems of natural doublet-triplet splitting (the second hierarchy
problem), the $\mu$ problem, and give predictions for proton decay,
neutrino masses and calculable threshold corrections at $M_G$.

\item On the bad side,  it is still not a fundamental theory and there
may not be a unique extension of the MESG to higher energies.
\end{itemize}

\begin{table}[t]
\begin{center}
\begin{tabular}{|c|c|c|}
\multicolumn{3}{l}{Table~4. }\\
 \multicolumn{3}{l}{}\\ \hline\hline
   & Good & Bad \\ \hline

   Eff. F.T.  & \underline{ 4 op's. at $M_G \Rightarrow$ {\bf LED}}&
\underline{Not fundamental} \\
  $\le M_G$ & 5 para's. $\Rightarrow$ 13 observables & \underline{No
symmetry}  \\
  & + 2 gauge para's. $\Rightarrow$ 3 observables & $\Rightarrow$ Why
these operators?  \\
    & \underline{+ 5} soft SUSY  &  (F.T. + Planck slop) \\

    &  breaking para's. $\Rightarrow \cdots$ &      \\

  &  Total {\bf 12} parameters &  \underline{Threshold corrections?} \\
\hline
Eff. F.T.  &  \underline{Symmetry} &  \underline{Not fundamental} \\
  $\le M $ & i)  gives Eff. F.T. $\le M_G$  &    \\
$M = M_{string}$ &         + corrections          &   \underline{Not
unique?} \\
 or $M_{Planck}$ & ii)  constrains other operators  &          \\
   &   \underline{GUT symmetry breaking}  &      \\
  &  i)  d - t splitting   &         \\
   &  ii)  $\mu$ problem   &          \\
    &  iii)  proton decay   &        \\
    &  iv)  neutrino masses  &        \\
     &  v)  threshold corrections at $M_G$  &     \\  \hline
\end{tabular}
\end{center}
\end{table}

\section{String Threshold at $M_S$}

Upon constructing the effective field theory $\le M_S$, we will have
determined the necessary SO(10) states, symmetries and couplings which
reproduce our fermion mass relations.  This theory can be the starting
point for constructing a realistic string model.  String model builders
could try to obtain a string vacuum with a massless spectra which
agrees with ours.  Of course, once the states are found the string will
determine the symmetries and couplings of the theory.   It is hoped
that in this way a {\it fundamental} theory of Nature can be found.
Work in this direction by several groups is in progress\cite{lykken}.  String
theories with SO(10), three families plus additional 16 + $\overline{16}$
pairs,   45's, 10's and even some 54 dimensional representations appear
possible.  One of the first results from this approach is the fact that only
one
of the three   families has diagonal couplings to the 10,  just as we have
assumed.

\section{Constructing the Effective Field Theory below $M_S$}

In this section I will discuss some preliminary results obtained in
collaboration with Lawrence Hall\cite{hr}. I will describe the necessary
ingredients for constructing model 6.  Some very general results from
this exercise are already apparent.

\begin{itemize}
\item {\it States} --- We have constructed a SUSY GUT which includes all
the states necessary for GUT symmetry breaking and also for generating
the 45 vevs in the desired directions.  A minimal representation
content below $M_S$ includes 54s + 45s + 3  16s + n($\overline{16}$ +
16) pairs  + 2  10s.
\item {\it Symmetry} --- In order to retain sufficient symmetry the
superspace potential in the visible sector W necessarily has a number
of flat directions.  In particular the scales $v_5$ and $v_{10}$ can
only be determined when soft SUSY breaking and quantum corrections are
included.  An auxiliary consequence is that the vev of W$_{visible}$
vanishes in the supersymmetric limit.
\item {\it Couplings} --- As an example of the new physics which results
from this analysis I will show how a solution to the $\mu$ problem, the
ratio $\lambda_b/\lambda_t$ and proton decay may be inter-related.
\end{itemize}

In Table 5 are presented the predictions for Model 6 for particular
values of the input parameters.

\begin{center}
\indent Table 5: Particular Predictions for Model 6
with $\alpha_s(M_Z) = 0.115$
\vskip 20pt
\begin{tabular}{|c|c|c|c|}
\hline
  Input  & Input & Predicted & Predicted  \cr
 Quantity &  Value &  Quantity & Value \cr
\hline
  $m_b(m_b)$ & $4.35 $ GeV & $M_t$ & $176$ GeV \cr
  $m_\tau(m_\tau)$ &$1.777 $ GeV & $\tan\beta$ & $55 $  \cr
\hline
  $m_c(m_c)$ &$1.22$ GeV & $V_{cb}$ & $.048 $ \cr
\hline
  $m_\mu$ &$105.6 $ MeV   & $V_{ub}/V_{cb}$ & $.059 $ \cr
  $m_e $   &$0.511$ MeV  &  $m_s(1GeV)$ & $172 $ MeV \cr
  $V_{us}$       &$0.221 $    & $\hat{B_K} $ & $0.64$  \cr
                 &            & $m_u / m_d$  & $0.64$  \cr
                 &            & $m_s / m_d$  & $24.$  \cr
\hline
\end{tabular}
\end{center}
In addition to these predictions, the set of inputs in
Table 5 predicts:

$\sin 2\alpha = -.46$, $\sin 2\beta = .49$,
$\sin 2\gamma = .84$, and $J = 2.6\times 10^{-5}$.

\begin{center}
{\Large\bf Model 6}
\end{center}

\vskip .1in

The superspace potential for Model 6 has several pieces -  W =
W$_{fermion}$ + W$_{symmetry \ breaking}$ + W$_{Higgs}$ +
W$_{neutrino}$.

\subsection{Fermion sector}

The first term must reproduce the four fermion mass operators of Model
6.  They are given by
\begin{eqnarray}
    O_{33} = &  16_3\ 10_1 \ 16_3 &  \nonumber \\
     O_{23} = & 16_2 \ {A_2 \over {\tilde A}} \ 10_1 \ {A_2 \over
{\tilde A}} \ 16_3 & \nonumber\\
    O_{22}  =  &  16_2 \ {{\tilde A} \over M} \ 10_1 \ {A_1 \over
{\tilde A}} \ 16_2 & \nonumber\\
    O_{12} = & 16_1 \left( {{\tilde A}\over M}\right)^3 \ 10_1 \left(
{{\tilde A} \over M} \right)^3 16_2 & \nonumber
\end{eqnarray}

There are two 10s in this model, denoted by $10_i, i= 1,2$ but only
$10_1$ couples to the ordinary fermions.   The A fields are different
45s which are assumed to have vevs in the following directions --
$\langle A_2\rangle = 45_Y$,  $\langle A_1\rangle = 45_{B-L}$, and  $\langle
\tilde A\rangle =   45_X$.  As noted earlier, there are 6 choices for the 22
operator and we have just chosen one of them, labelled a, arbitrarily here.
In
figure 20,  we give the tree diagrams which reproduce the effective
operators for Model 6 to leading order in an expansion in the ratio of
small to large scales.  The states  $\overline{\Psi}_a, \Psi_a, a = 1,
\cdots ,9$ are massive $\overline{16}, 16$ states respectively with mass given
by $\langle {\cal S}_M \rangle \sim M$.   Each
vertex represents a separate Yukawa interaction in W$_{fermion}$ (see
below).  Field theoretic corrections to the effective GUT operators may
be obtained by diagonalizing the mass matrices for the heavy states and
integrating them out of the theory.

\begin{eqnarray}
W_{fermion} =  &    &  \nonumber
\end{eqnarray}

$$ 16_3 16_3 10_1  +  \opsi_1 A_2 16_3  +  \opsi_1 {\tilde A} \Psi_1  +  \Psi_1
\Psi_2 10_1  $$

$$ +  \opsi_2 {\tilde A} \Psi_2  +  \opsi_2 A_2 16_2  +  \opsi_3  A_1 16_2  $$

$$ +  \opsi_3 {\tilde A} \Psi_3  +  \Psi_3 \Psi_4 10_1  +  {\cal S}_M
\sum_{a=4}^9  ( \opsi_a \Psi_a )  $$

$$ + \opsi_4 {\tilde A} 16_2  +  \opsi_5 {\tilde A} \Psi_4
+  \opsi_6 {\tilde A} \Psi_5  $$

$$  +  \Psi_6 \Psi_7 10_1  +  \opsi_7 {\tilde A} \Psi_8  +  \opsi_8 {\tilde A}
\Psi_9  +  \opsi_9 {\tilde A} 16_1 $$

\begin{itemize}
\item {\em Note that the vacuum insertions in the effective operators
above cannot be rearranged,  otherwise an inequivalent low energy
theory would result.  In order to preserve this order naturally we
demand that each field carries a different value of a U(1) family
charge (see fig. 20).} Note also that the particular choice of a 22
operator will affect the allowed U(1) charges of the states.  Some
choices may be acceptable and others not.
\item Consider W$_{fermion}$.  It has many terms,  each of which can
have different, in principle, complex Yukawa couplings.  {\em
Nevertheless the theory is predictive because only a very special
linear combination of these parameters enters into the effective theory
at $M_G$.  Thus the observable low energy world is simple, not because
the full theory is particularly simple, but because the symmetries are
such that the effective low energy theory contains only a few dominant
terms.}
\end{itemize}

\subsection{Symmetry breaking sector}

The symmetry breaking sector of the theory is not particularly
illuminating.  Two 54 dimensional representations, $S, S'$ are needed
plus several singlets denoted by ${\cal S}_i, i = 1, \cdots, 7$.  They
appear in the first two terms and are responsible for driving the vev
of $A_1$ into the B-L direction,  the third term drives the vev of the
$\overline{16},16$ fields $\overline{\Psi}, \Psi$ into the right-handed
neutrino like direction breaking SO(10) to SU(5) and forcing ${\tilde
A}$ into the X direction.  The fourth, fifth and sixth terms drive $A_2$ into
the Y direction. Finally the last two terms are necessary in order to
assure that all non singlet states under the SM gauge interactions
obtain mass of order the GUT scale.  All  primed fields are assumed to
have vanishing vevs.

Note if $\langle {\cal S}_3\rangle \approx M_S$ then two of these adjoints
state
may be heavy.  Considerations such as this will affect how couplings
run above $M_G$.

\begin{eqnarray}
W_{symmetry \,\, breaking} =  &    &  \nonumber
\end{eqnarray}

$$    A'_1 (S A_1 + {\cal S}_1 A_1 ) +  S' ( {\cal S}_2 S + A_1^2) $$

$$  + {\tilde A}' ( \opsi \Psi + {\cal S}_3  \tilde A ) $$

$$ +  A'_2 ( {\cal S}_4 A_2  +  S  {\tilde A} + ( {\cal S}_1 + {\cal S}_5)
{\tilde A} ) $$

$$ +   \opsi' A_2 \Psi  +  \opsi A_2 \Psi' $$

$$ +  A_1 A_2 {\tilde A}'  +  {\cal S}_6 (A'_1)^2  $$

\subsection{Higgs sector}

The Higgs sector is introduced below.  It does not at the moment appear
to be  unique, but it is crucial for understanding the solution to
several important problems -- doublet-triplet splitting,  $\mu$ problem
and proton decay -- and these constraints may only have one solution.
The $10_1 A_1 10_2$ coupling is  the term required by the
Dimopoulos-Wilczek mechanism for doublet-triplet splitting.  Since
$A_1$ is an anti-symmetric tensor, we need at least two 10s.

The couplings of $10_1$ to the 16s are introduced to solve the $\mu$
problem.  After naturally solving the doublet-triplet splitting problem
one has massless doublets.  One needs however a small supersymmetric
mass $\mu$ for the Higgs doublets of order the weak scale.  This may be
induced once SUSY is broken in several ways.

\begin{itemize}
\item The vev of the field $A_1$ may shift by an amount of order the
weak scale due to the introduction of the soft SUSY breaking terms into
the potential. In this theory the shift of $A_1$ appears to be too
small.
\item There may be higher dimension D terms in the theory of the form,  eg.
$${1 \over M_{Pl}}\int d^4\theta 10_1^2 (A_2^*).  $$  Then supergravity
effects might induce a non-vanishing vev for the F term of $A_2$ of
order the $m_W M_G$.  This will induce a value of $\mu$ of order  $m_W
M_G/M_{Pl}$.  The shift in the F-terms also appear to be negligible.
\item Higher dimension D-terms with hidden sector fields may however work.
Consider ${1 \over M_{Pl}} \int d^4\theta 10_1^2 z^*$  where $z$ is a hidden
sector field which is connected with soft SUSY breaking.  It would then be
natural to have  $F_z \approx \mu M_{Pl}$.
\item One loop effects may induce a $\mu$ term once soft SUSY breaking
terms are introduced\cite{hall}.  In this case we find  $\mu \sim  {A
\lambda^4 \over 16 \pi^2}$  where $\lambda^4$ represents the product of
Yukawa couplings entering into the graph of figure 21.

We use the last mechanism above for generating $\mu$ in the example which
follows.

\end{itemize}

\begin{eqnarray}
W_{Higgs} =  &    &  \nonumber
\end{eqnarray}

$$ +   \opsi' A_2 \Psi  +  \opsi A_2 \Psi' $$

$$ +   10_1  A_1 10_2  +  {\cal S}_7 10_2^2  $$

$$ +   \opsi \opsi' 10_1  +  \Psi \Psi' 10_1   $$

Note that the first two terms already appeared in the discussion of the
symmetry breaking sector.  They are included again here since as you
will see they are important for the discussion of the Higgs sector as well.
The last two terms are needed to incorporate the solution to the $\mu$ problem.
As a result of these couplings to $\overline{\Psi}, \Psi$ the
Higgs doublets in $10_1$  mix with other states.  The mass matrix for
the SU(5) ${\bf \overline{5}, 5}$ states in ${\bf 10_1, 10_2,\Psi,
\Psi', \overline{\Psi}, \overline{\Psi}'}$ is given below.

 \begin{eqnarray}  &\overline{5}_1   ~\overline{5}_2
{}~\overline{5}_{\Psi}  ~\overline{5}_{\Psi'} & \nonumber \\
\begin{array}{c} 5_1 \\ 5_2 \\ 5_{\overline{\Psi}} \\
5_{\overline{\Psi}'}
 \nonumber \end{array} &
\left( \begin{array}{cccc}  0 & A_1 & 0 & \Psi \\
                            A_1 & {\cal S}_7 & 0 & 0 \\
                            0 &  0 & 0 & A_2 \\
                            \overline{\Psi} & 0 & A_2 & 0 \end{array}
\right)& \nonumber
\end{eqnarray}

{\bf Higgs doublets} In the doublet sector the vev $A_1$ vanishes.
Since the Higgs doublets in 10$_1$ now mix with other states, the
boundary condition $\lambda_b/\lambda_t = 1$ is corrected at tree
level.  The ratio is now given in terms of a ratio of mixing angles.

{\bf Proton decay} The rate for proton decay in this model is set by
the quantity  $(M^t)^{-1}_{11}$ where $M^t$ is the color triplet
Higgsino mass matrix\cite{drw2}.  We find  $(M^t)^{-1}_{11}= {{\cal S}_7 \over
A_1^2}$.  This may be much smaller than ${1 \over M_G}$ for ${\cal
S}_7$ sufficiently smaller than $M_G$.  Note there are no light color
triplet states in this limit.  Proton decay is suppressed since in this
limit the color triplet Higgsinos in $10_1$ become Dirac fermions (with
mass of order $M_G$), {\em but they do not mix with each other}.

\subsection{Symmetries} The theory has been constructed in order to
have enough symmetry to restrict the allowed operators.  This is
necessary in order to reproduce the mass operators in the effective
theory, as well as to preserve the vacuum directions assumed for the
45s and have natural doublet-triplet splitting.  Indeed the
construction of the symmetry breaking sector with the primed fields
allows the 45s to carry nontrivial U(1) charges.   This model has
several unbroken U(1) symmetries which do not seem to allow any new
mass operators.  It has a discrete $Z_4$ R parity in which all the
primed fields, ${\cal S}_{6,7}$ and $10_2$ are odd and $16_i, i =
1,2,3$ and $\overline{\Psi}_a, \Psi_a, a = 1, \cdots , 9$ go into $i$
times themselves. This guarantees that the odd states (and in
particular, $10_2$) do not couple into the fermion mass sector.  There
is in addition a Family Reflection Symmetry (see Dimopoulos- Georgi
\cite{drw1}) which guarantees that the lightest supersymmetric particle
is stable.  Finally, there is a continuous R symmetry which is useful for two
reasons, (1)  as a consequence,  only dimension 4 operators appear in the
superpotential and (2) this R symmetry is an anomalous Peccei-Quinn U(1) which
naturally solves the strong CP problem.

{\bf Neutrino sector}  The neutrino sector seems to be very model
dependent.  It will constrain the symmetries of the theory, but I will
not discuss it further here.

\section{Conclusion}

In this last lecture, I have presented a class of supersymmetric SO(10) GUTs
which are  in {\em quantitative} agreement with the low energy data.  With
improved data these particular models may eventually be ruled out. Nevertheless
the approach of using low energy data to ascertain the dominant operator
contributions at $M_G$ seems robust.  Taking it seriously, with {\em
quantitative} fits to the data and including the leading order corrections to
the zeroth order results, may eventually   lead us to the correct theory.

What is the proverbial {\em smoking gun} for the theories presented here ?
There are three observations which combined would confirm SUSY GUTs.
\begin{enumerate}
\item Gauge coupling unification consistent with the observed values of
$\alpha,
 sin^2\theta_W, \alpha_s$.
\item Observation of SUSY particles.
\item Observation of proton decay into the modes
$p \rightarrow K^+ \overline{\nu}$ and $p \rightarrow K^0 \mu^+$~\cite{drw2}.
Although SUSY GUTs may not predict the rate for
this process, nevertheless the observation of this process would confirm SUSY
GUTs.
\end{enumerate}

In addition,
the minimal SO(10) models presented here all demand large $\tan\beta$.  Thus
observation of large $\tan\beta$ would certainly strengthen these ideas.
Finally, if the calculable corrections to the predictions of one of these
models
improve the agreement with the data, it would be difficult not to accept this
theory as a true description of nature.

\section{Acknowledgements}

I would like to thank all my collaborators with whom I have discussed much of
the
material in these lectures.  I would also like to thank the organizers of the
summer school for the enjoyable week I spent in Trieste.  This work is
supported in part by U.S. Department of Energy contract DE-ER-01545-640.

\end{document}